\documentclass[12pt]{article}
\usepackage{graphicx}
\usepackage{natbib} 
\usepackage{url} 
\usepackage{amsmath} 
\usepackage{amsfonts}
\usepackage{amssymb} 
\usepackage{lineno}
\usepackage{setspace}
\usepackage{lscape} 
\usepackage{float} 
\usepackage{breakcites}
\usepackage{pdfpages}
\usepackage{bm}
\usepackage{xcolor}
\usepackage{hyperref}
\usepackage{setspace}

\graphicspath{{Figures/}}

\newenvironment{nalign}{
    \begin{equation}
    \begin{aligned}
}{
    \end{aligned}
    \end{equation}
    \ignorespacesafterend
}

\usepackage{mathtools}

\newcommand{\interior}[1]{%
 {\kern0pt#1}^{\mathrm{o}}%
}
\usepackage [english]{babel}
\usepackage [autostyle, english = american]{csquotes}
\MakeOuterQuote{"}

\usepackage{algorithm,algpseudocode}
\usepackage{caption}

\usepackage[makeroom]{cancel}

\newcommand{\mbR}{{\mathbb R}}

\newcommand{\blind}{0}

\begin{document}

\def\spacingset#1{\renewcommand{\baselinestretch}%
{#1}\small\normalsize} \spacingset{1}

\if0\blind
{
  \title{\bf A Scalable Partitioned Approach to Model Massive Nonstationary Non-Gaussian Spatial Datasets}
 \author{Benjamin Seiyon Lee and Jaewoo Park\thanks{Corresponding Author: Department of Statistics and Data Science, Yonsei University; Department of Applied Statistics, Yonsei University, Seoul, 03722, Republic of Korea (e-mail: jwpark88@yonsei.ac.kr)}\hspace{.2cm}\\
Department of Statistics, George Mason University \hspace{.2cm}\\
Department of Statistics and Data Science, Yonsei University \hspace{.2cm}\\
Department of Applied Statistics, Yonsei University}
  \maketitle
} \fi

\if1\blind
{
  \bigskip
  \bigskip
  \bigskip
  \begin{center}
    {\LARGE\bf A Scalable Partitioned Approach to Model Massive Nonstationary Non-Gaussian Spatial Datasets}
\end{center}
  \medskip
} \fi

\begin{abstract}
Nonstationary non-Gaussian spatial data are common in many disciplines, including climate science, ecology, epidemiology, and social sciences. Examples include count data on disease incidence and binary satellite data on cloud mask (cloud/no-cloud). Modeling such datasets as stationary spatial processes can be unrealistic since they are collected over large heterogeneous domains (i.e., spatial behavior differs across subregions). Although several approaches have been developed for nonstationary spatial models, these have focused primarily on Gaussian responses. In addition, fitting nonstationary models for large non-Gaussian datasets is computationally prohibitive. To address these challenges, we propose a scalable algorithm for modeling such data by leveraging parallel computing in modern high-performance computing systems. We partition the spatial domain into disjoint subregions and fit locally nonstationary models using a carefully curated set of spatial basis functions. Then, we combine the local processes using a novel neighbor-based weighting scheme. Our approach scales well to massive datasets (e.g., 1 million samples) and can be implemented in \texttt{nimble}, a popular software environment for Bayesian hierarchical modeling. We demonstrate our method to simulated examples and two large real-world datasets pertaining to infectious diseases and remote sensing.






\end{abstract}
\noindent%
\vfill

\newpage
\spacingset{2} 

\section{Introduction}

Nonstationary spatial models have been used in a wide range of scientific applications, including disease modeling \citep{ejigu2020geostatistical}, remote sensing \citep{heaton2017nonstationary}, precision agriculture \citep{katzfuss2013bayesian}, precipitation modeling \citep{risser2015local}, and air pollutant monitoring \citep{fuentes2002interpolation}. Simple models assume second-order stationarity of the spatial process; however, this can be unrealistic since data are collected over heterogeneous domains. Here, the spatial processes can exhibit localized spatial behaviors. Although several methods \citep[cf.][]{fuentes2002interpolation, risser2015local, heaton2017nonstationary} have been developed for modeling nonstationary spatial data, these have focused primarily on Gaussian spatial data. Moreover, fitting these models poses both computational and inferential challenges, especially for large datasets. In this manuscript, we propose a scalable algorithm for fitting nonstationary non-Gaussian datasets. Our approach captures nonstationarity by partitioning the spatial domain and modeling local spatial processes using basis expansions. This new algorithm is computationally efficient in that: (1) partitioning the spatial domain permits parallelized computation on high-performance computation (HPC) systems; and (2) basis approximation of spatial processes dramatically reduces the computational overhead. 

There is a growing literature on modeling nonstationary spatial datasets. Weighted average methods \citep{fuentes2001high, kim2005analyzing, risser2015local} combine localized spatial models to reduce computational costs. Basis function approximations \citep{nychka2002multiresolution, katzfuss2013bayesian, katzfuss2017multi,hefley2017basis} represent complex spatial processes using linear combinations of spatial basis functions. \cite{higdon1998process} and \cite{paciorek2006spatial} represent a nonstationary process using convolutions of spatially varying kernel functions. Based on the spatial partitioning strategies, some of these approaches are amenable to massive spatial datasets. For example, \cite{heaton2017nonstationary} develops a computationally efficient approach for large nonstationary spatial data by partitioning an entire domain into disjoint sets using a hierarchical clustering algorithm. \cite{katzfuss2017multi} constructs basis functions at multiple levels of resolution based on recursive partitioning of the spatial region. \cite{guhaniyogi2018meta} proposes a divide-and-conquer approach to generate a global posterior distribution by combining local posterior distributions from each subsample. Though these approaches scale well, they are limited to Gaussian responses. 

Spatial generalized linear mixed models (SGLMMs) \citep{diggle1998model} are popular class of models designed for non-Gaussian spatial datasets. SGLMMs are widely used for both areal and point-referenced data, where latent Gaussian random fields can account for the spatial correlations. However, fitting SGLMMs for massive spatial datasets is computationally demanding since the dimension of correlated spatial processes grows with an increasing number of observations. Although several computational methods \citep{banerjee2008gaussian, hughes2013dimension, guan2018computationally, lee2019picar, zilber2020vecchia} have been proposed for large non-Gaussian spatial datasets, these methods assume second-order stationarity of the latent spatial processes.


In this manuscript, we propose a scalable approach for modeling massive nonstationary non-Gaussian spatial datasets. Our smooth mosaic basis approximation for nonstationary SGLMMs (SMB-SGLMMs) combines key ideas from weighted average approaches and basis approximations. SMB-SGLMM consists of four steps: (1) partition the spatial region using a spatial clustering algorithm \citep{heaton2017nonstationary}; (2) generate localized spatial basis functions; (3) fit a nonstationary basis-representation model to each partition; and (4) smooth the local processes using distance-based weighting scheme (smooth mosaic). Due to the partitioning and localized model fitting, we can leverage parallel computing, which greatly increases the scalability of the SMB-SGLMM method. To our knowledge, this study is the first attempt to develop a scalable algorithm for fitting large nonstationary non-Gaussian spatial datasets. Furthermore, our method provides an automated mechanism for selecting appropriate spatial basis functions. We also provide ready-to-use code written in \texttt{nimble} \citep{nimble2017}, a software environment for Bayesian inference.

The outline for the remainder of this paper is as follows. In Section 2, we introduce several nonstationary modeling approaches. We discuss the potential extension of stationary SGLMMs to nonstationary SGLMMs and discuss their challenges. In Section 3, we propose SMB-SGLMMs for massive spatial data and provide implementation details. Furthermore, we investigate the computational complexity of our method in detail. In Section 4, we study the performance of SMB-SGLMMs through simulated data examples. In Section 5, we apply SMB-SGLMMs to malaria incidence data and binary cloud mask data from satellite imagery. We conclude with a discussion and summary in Section 6.

\section{Nonstationary Modeling for Non-Gaussian Spatial Data}

Let $\mathbf{Z}=\lbrace Z(\mathbf{s}_i) \rbrace_{i=1}^{N}$ be the observed data and $\mathbf{X} \subset \mbR^{N\times p}$ be the matrix of covariates at the spatial locations $\mathbf{s}=\lbrace \mathbf{s}_i\rbrace_{i=1}^{N}$ in a spatial domain $\mathcal{S} \subseteq \mbR^2$. $\mathbf{W}=\lbrace W(\mathbf{s}_i) \rbrace_{i=1}^{N}$ is a mean-zero Gaussian process with covariance matrix $\Sigma\subset \mbR^{N\times N}$. Then SGLMMs can be defined as 
\begin{nalign} 
g\lbrace \mathbb{E}[\mathbf{Z}|\bm{\beta}, \mathbf{W}] \rbrace:= \bm{\eta} & = \mathbf{X}\bm{\beta} + \mathbf{W}\\
\mathbf{W} & \sim N(0,\Sigma)
\label{Model}
\end{nalign}
with link function $g(\cdot)$ and linear predictor $\bm{\eta}$. Standard SGLMMs \citep{diggle1998model} consider a second-order stationary Gaussian process for $\mathbf{W}$ for their convenient mathematical framework. However, this assumption can be unrealistic for spatial processes existing in large heterogeneous domains (see \cite{bradley2016comparison}, for a discussion). A natural extension to \eqref{Model} is to model $\mathbf{W}$ as a nonstationary spatial process. There is an extensive literature on modeling nonstationary spatial data \citep{sampson2010constructions} such as: (1) weighted-average methods, (2) basis function methods, and (3) process convolutions. Our method is motivated by these nonstationary modeling approaches. 

Weighted average methods \citep{fuentes2001high} divide the spatial region $\mathcal{S}$ into disjoint partitions and fit locally stationary models to each partition. For example, \cite{kim2005analyzing, heaton2017nonstationary} partition the spatial domain through Voronoi tessellation. Then, the global process is constructed by combining the locally stationary processes via a weighted average. The weights are computed using the distances between the observation locations and `center' of the localized processes. These approaches scale well by taking advantage of parallel computation \citep[cf.][]{risser2015local, heaton2017nonstationary}.

Basis functions approaches represent the nonstationary covariance structure as an expansion of spatial basis functions $\lbrace \Phi_{j}(\mathbf{s}) \rbrace_{j=1}^{m}$. Let $\bm{\Phi}$ be an $N$ by $m$ matrix with columns indicate the basis functions and rows indicate locations $\bm{\Phi}_{i,j}=\Phi_{j}(\mathbf{s}_{i})$. Then we can construct a nonstationary spatial process as 
\[
\mathbf{W} \approx \bm{\Phi}\bm{\delta},~~~~\bm{\delta} \sim N(0,\bm{\Sigma}_{\bm{\Phi}}),
\]
where $\bm{\delta}$ is the coefficients of basis functions. We approximate the covariance structure as $\bm{\Phi}\bm{\Sigma}_{\bm{\Phi}}\bm{\Phi}^{\top}$, which is not dependent solely on the lag between locations; hence this is nonstationary. Different types of basis functions have been used, for instance eigenfunctions obtained from the empirical covariance \citep{holland1999spatial}, multiresolution basis functions \citep{nychka2002multiresolution,nychka2015multiresolution,katzfuss2017multi}, and computationally efficient low-rank representation of nonstationary covariance \citep{katzfuss2013bayesian}.

Process convolutions represent the nonstationary spatial processes through convolutions of spatially varying kernel function and Brownian motion. For an arbitrary $\mathbf{s} \in \mathcal{S}$,
\[
\mathbf{W}(\mathbf{s}) = \int_{\mathcal{S}}
K_{\mathbf{s}}(\mathbf{u})d\mathbf{W}(\mathbf{u})
\]
where $K_{\mathbf{s}}(\cdot)$ is a kernel function centered at location $\mathbf{s}$ and $\mathbf{W}(\cdot)$ is a bivariate Brownian motion. \cite{higdon1998process} use bivariate Gaussian kernels under this framework. Several extensions have also been proposed including creating closed-form nonstationary Mat\'{e}rn covariance functions \citep{paciorek2006spatial}, extension to multivariate spatial process \citep{kleiber2012nonstationary}, and computationally efficient local likelihood approaches \citep{risser2015local}.

We note that these nonstationary models have focused on Gaussian responses. Direct application of these methods to \eqref{Model} is challenging because we cannot obtain closed-form maximum likelihood estimates by marginalizing out $\mathbf{W}$. Within the Bayesian framework, updating conditional posterior distributions requires a computational complexity of $\mathcal{O}(N^3)$, which becomes infeasible even for moderately large size datasets (e.g., binary satellite data with 100,000 observations). Although several computationally efficient approaches \citep[cf.][]{rue2009approximate,hughes2013dimension, guan2018computationally,lee2019picar, zilber2020vecchia} have been developed for non-Gaussian hierarchical spatial models, they are assuming stationarity of $\mathbf{W}$. In what follows, we develop partitioned nonstationary models for non-Gaussian spatial data. Our method is computationally efficient and provides accurate predictions over large heterogeneous spatial domains.

\section{Smooth Mosaic Basis Approximation for Nonstationary SGLMMs}

We propose a smooth mosaic basis approximation for nonstationary SGLMMs (SMB-SGLMMs) designed for massive spatial datasets. We begin with an outline of our method: \\

\noindent {\it{Step 1.}} Partition the spatial domain into disjoint subregions.\\
\noindent {\it{Step 2.}} Construct data-driven basis functions for each subregion.\\
{\it{Step 3.}} Fit a locally nonstationary basis function model to each subregion in parallel. \\
{\it{Step 4.}} Construct the global nonstationary process as a weighted average of local processes.\\

\begin{figure}
\begin{center}
\includegraphics[ scale = 0.9]{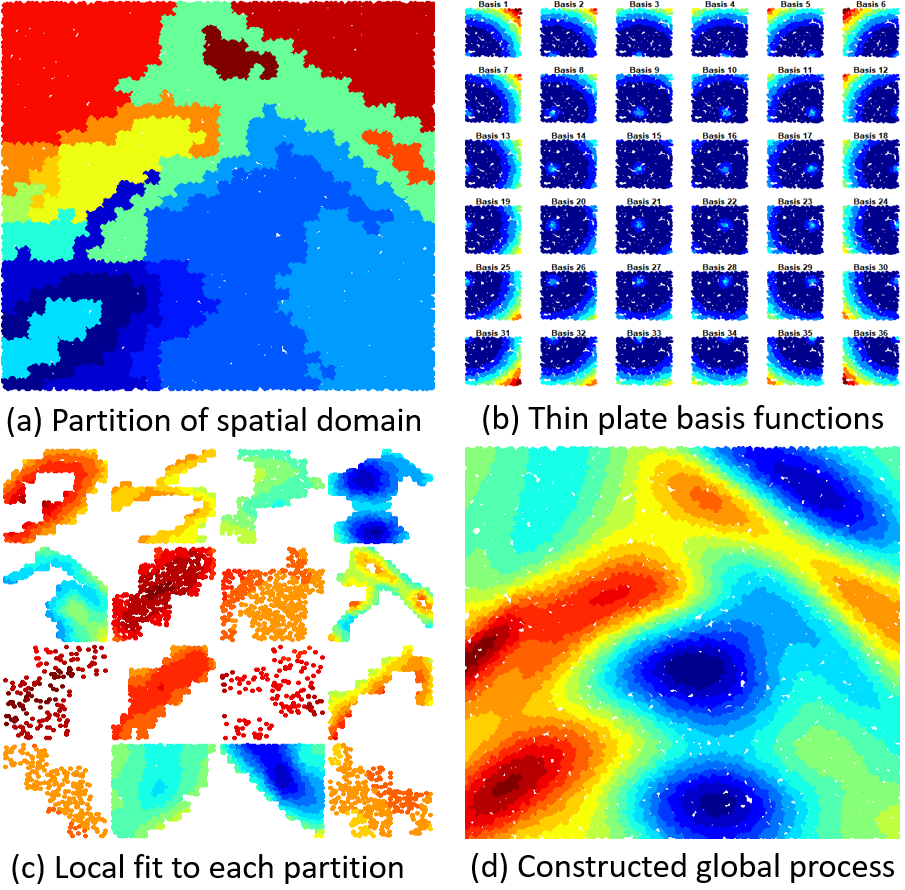}
\end{center}
\caption[]{Illustration for the partitioned nonstationary approach for simulated $\mathbf{W}$. (a) Nonstationary $\mathbf{W}$ is partitioned through 16 subregions; different colors indicate disjoint partitions. (b) For each partition, thin plate splines basis functions are constructed at knots; basis functions represent distinct spatial patterns. (c) The Local nonstationary model is fit to each partition using a linear combination of basis functions. (d) The global nonstationary process is obtained via a weighted average of the local processes.}
\label{summary}
\end{figure}

\noindent SMB-SGLMMs are described in Figure~\ref{summary}. We provide the details in the following subsections. 

\subsection{Partitioned Nonstationary Spatial Models}\label{SubSec:partition}


\subsubsection*{Step 1. Partition the spatial domain into disjoint subregions}

We use an agglomerative clustering approach \citep{heaton2017nonstationary} to partition the spatial domain $\mathcal{S}$ into $K$ subregions $\lbrace \mathcal{S}_k \rbrace_{k=1}^{K}$, which satisfy $\cup_{k=1}^{K} \mathcal{S}_k = \mathcal{S}$. We fit a nonspatial generalized linear model ({\tt{glm}} function in {\tt{R}}) using responses $\mathbf{Z}$ and covariates $\mathbf{X}$. Then we obtain the spatially correlated residuals $\lbrace \epsilon(\mathbf{s}_i) \rbrace_{i=1}^{N}$. For all $i\neq j$, we calculate the dissimilarity between $\epsilon(\mathbf{s}_i)$ and $\epsilon(\mathbf{s}_i)$ as $d_{ij} = |\epsilon(\mathbf{s}_i)-\epsilon(\mathbf{s}_j)|/\|\mathbf{s}_i-\mathbf{s}_j\|$ from spatial finite differences \citep{banerjee2006bayesian}. \cite{heaton2017nonstationary} assigns locations with low dissimilarity values ($d_{ij}$) into the same partitions. The main idea is to separate locations with large pairwise dissimilarities (i.e. rapidly changing residual surfaces $\epsilon(\mathbf{s})$). We initialize $K=N$ where each observation belongs to its own cluster. Then we combine two clusters if they are Voronoi neighbors and have minimum pairwise dissimilarity. We repeat this procedure until we arrive at the desired $K$ partitions (Figure~\ref{summary} (a)). We provide details about the clustering algorithm in the supplementary material. 

\subsubsection*{Step 2. Construct data-driven basis functions for each subregion}

For each partition, we generate a collection of spatial basis functions. We have $\mathbf{Z}_{k}=\lbrace Z(\mathbf{s}): \mathbf{s} \in \mathcal{S}_k \rbrace \in \mbR^{N_{k}}$, the observations belong to $\mathcal{S}_k$, where $N=\sum_{k=1}^{K}N_k$. $\mathbf{X}_k$ is an $N_k \times p$ matrix of covariates. Consider the knots (grid points) $\lbrace \mathbf{u}_{kj} \rbrace_{j=1}^{m_k}$ over $\mathcal{S}_k$ ($m_k \ll N_k$). These knots can define a wide array of spatial basis functions such as radial basis functions \citep{nychka2015multiresolution,katzfuss2017multi} and eigenbasis functions \citep{banerjee2008gaussian}. In this study, we consider thin plate splines defined as $\Phi_{kj}(\mathbf{s})=\|\mathbf{s}-\mathbf{u}_{kj}\|^2 \log(\|\mathbf{s}-\mathbf{u}_{kj}\|)$. Here $\bm{\Phi}_k$ is an $N_k \times m_k$ matrix by evaluating the basis function at $N_k$ locations in $\mathcal{S}_k$ (Figure~\ref{summary} (b)). Although we focus on thin plate splines, different types of basis functions can be considered. Examples include eigenfunctions \citep{holland1999spatial,Banerjee2013,guan2018computationally}, radial basis \citep{nychka2015multiresolution,katzfuss2017multi}, principal components \citep{higdon2008computer,cressie2015statistics}, and Moran's basis \citep{hughes2013dimension, lee2019picar}.


\subsubsection*{Step 3. Fit a locally nonstationary basis function model to each subregion in parallel.}

For each partition, we can represent the spatial random effects as $\mathbf{W}_k \approx \bm{\Phi}_k\bm{\delta}_k$ and model the conditional mean $\mathbb{E}[\mathbf{Z}_k|\bm{\beta}_k, \bm{\Phi}_k, \bm{\delta}_k]$ as   
\begin{nalign} 
g\lbrace \mathbb{E}[\mathbf{Z}_k|\bm{\beta}_k, \bm{\Phi}_k, \bm{\delta}_k] \rbrace & := \bm{\eta}_k = \mathbf{X}_k\bm{\beta}_k + \bm{\Phi}_k\bm{\delta}_k\\
\bm{\delta}_k & \sim N(0,\bm{\Sigma}_{\bm{\Phi}_k}),
\label{partitionedModel}
\end{nalign}
where $\bm{\Sigma}_{\bm{\Phi}_k}$ is a covariance of basis coefficients $\bm{\delta}_k$. Here we set $\bm{\Sigma}_{\bm{\Phi}_k}=\sigma^{2}_{k}\mathbf{I}_k$, as in a discrete approximation of a nontationary Gaussian process \citep{higdon1998process}. This basis representation approximates the covariance through $\sigma^{2}_{k}\bm{\Phi}_k\bm{\Phi}_{k}^{\top}$. Such approximation can capture the nonstationary behavior of the spatial process through a linear combination of basis functions (Figure~\ref{summary} (c)). Since we typically choose $m_k \ll N_k$, basis representations can drastically reduce computational costs by avoiding large matrix operations. For our simulated example (Section \ref{SubSec:SimulatedCount}), we use $m_k=81$ for a partition of size $N_k=13,129$. We provide implementation details in Section~\ref{SubSec:tune}. In addition, a clever choice of $\bm{\Phi}_k$ can also reduce correlations in $\bm{\delta}_k$, resulting in fast mixing MCMC algorithms \citep{haran2003accelerating,Christensen2006robust}. For the exponential family distribution $f(\cdot)$, the partition-specific hierarchical spatial model is as follows:
\begin{equation}
\begin{split}
& \textbf{Data Model: } \phantom{AAAAA} \mathbf{Z}_{k}|\bm{\eta}_k \sim f(\bm{\eta}_k)\\
\phantom{A}& \phantom{AAAAAAAAAAAAAA}g\lbrace \mathbb{E}[\mathbf{Z}_k|\beta_k,\bm{\Phi}_k,\bm{\delta}_k)] \rbrace:= \bm{\eta}_k =\mathbf{X}_k\beta_k + \Phi_k\bm{\delta}_k\\
& \textbf{Process  Model: } \phantom{AAA} \bm{\delta}_k|\sigma^{2}_{k} \sim N(0,\sigma^{2}_{k}\mathbf{I}_k)\\
& \textbf{Parameter  Model: } \phantom{A} \beta_k \sim p(\beta_k), \sigma^{2}_{k}\sim p(\sigma^{2}_{k})
\end{split}
\label{LocationHierarchy}
\end{equation}
We complete the hierarchical model by assigning prior distributions for the model parameters $\beta_k$ and $\sigma^{2}_{k}$. 


\subsubsection*{Step 4. Construct the global nonstationary process as a weighted average of the local processes.}

To construct the global process, we use a weighted average of the fitted local processes. 
Note that $\bm{\Phi}_k\in \mbR^{N_k\times m_k}$ is the basis functions matrix consisting of thin plate splines $\Phi_{kj}(\mathbf{s})=\|\mathbf{s}-\mathbf{u}_{kj}\|^2 \log(\|\mathbf{s}-\mathbf{u}_{kj}\|)$ for $s \in \mathcal{S}_k$, where $\lbrace \mathbf{u}_{kj} \rbrace_{j=1}^{m_k}$ are the knots over $\mathcal{S}_k$. Here, we introduce another notation. We define $\bm{\widetilde{\Phi}}_k\in \mbR^{N\times m_k}$ by evaluating $\Phi_{kj}(\mathbf{s})$ for all $s \in \mathcal{S}$. Let $\bm{\widetilde{\Phi}}_k(\mathbf{s}) \in \mbR^{m_k}$ be the row of $\bm{\widetilde{\Phi}}_k$ corresponding to spatial location $\mathbf{s}\in \mathcal{S}$.
Since $\mathbf{W_k(\mathbf{s})}\approx \bm{\widetilde{\Phi}}^{\top}_k(\mathbf{s})\bm{\delta}_k$, we have:
\begin{nalign} 
\mathbf{W(\mathbf{s})} & = \sum_{k=1}^{K} c_{k}(\mathbf{s})\mathbf{W_k(\mathbf{s})} \approx \sum_{k=1}^{K}c_{k}(\mathbf{s})\bm{\widetilde{\Phi}}^{\top}_k(\mathbf{s})\bm{\delta}_k,\\
c_k(\mathbf{s}) & \propto \exp\Big(-\|\mathbf{s}-\widetilde{\mathbf{s}}_{k}\|^2 \Big)~~\mbox{if}~\|\mathbf{s}-\widetilde{\mathbf{s}}_k\| \leq \gamma~~\mbox{otherwise}~0,
\label{globalW}    
\end{nalign}
where $\widetilde{\mathbf{s}}_{k} \in \mathcal{S}_k$ is the closest point to $\mathbf{s}$.
The weight $c_k(\mathbf{s})$ is proportional to the inverse distance between $\mathbf{s}$ and $\widetilde{\mathbf{s}}_k$; hence, shorter distances result in higher weights. We assign a 0 weight if the distance exceeds a threshold, or weighting radius, $\gamma$. We present details about choice of $\gamma$ in Section~\ref{SubSec:tune}. The weighted average of the local processes approximates the nonstationary global process (Figure~\ref{summary} (d)). Similarly, a global linear predictor $\bm{\eta}$ can be written as 
\begin{equation}
\bm{\eta(\mathbf{s})}  =\sum_{k=1}^{K}[\mathbf{X}(\mathbf{s})^{\top}\bm{\beta}_k\mathbf{1}_{\{\mathbf{s}\in \mathcal{S}_k\}} + c_{k}(\mathbf{s})\bm{\widetilde{\Phi}}^{\top}_k(\mathbf{s})\bm{\delta}_k],
\label{globaleta}
\end{equation}
where $\mathbf{1}_{\{\mathbf{s}\in \mathcal{S}_k\}}$ is an indicator function. Here $\mathbf{X}(s) \in \mbR^{p}$ is a vector of the covariate matrix $\mathbf{X}$ for location $\mathbf{s}$, $\bm{\beta}_k \in \mbR^{p}$ is corresponding regression coefficients. Our method provides a partition varying estimate of $\bm{\beta}_k$. This is because the fixed effects may have spatially varying (nonstationary) behavior over large heterogeneous spatial domains. Therefore, as in \cite{heaton2017nonstationary} we provide a partition varying $\bm{\beta}_k$ in our applications. If estimating a global $\bm{\beta}$ is of interest, one may consider divide and conquer algorithms such as consensus Monte Carlo \citep{scott2016bayes} or geometric median of the subset posteriors \citep{minsker2017robust}. Such methods provide the global posterior distribution of fixed effects by combining subset posteriors. 

\subsubsection*{Spatial prediction}

Spatial prediction at unobserved locations is of great interest in many scientific applications. Let $\mathbf{s}^{\ast} \in \mathcal{S}$ be an arbitrary unobserved location. From thin plate splines basis functions, we can construct a local basis as $\Phi_{kj}(\mathbf{s}^{\ast})=\|\mathbf{s}^{\ast}-\mathbf{u}_{kj}\|^2 \log(\|\mathbf{s}^{\ast}-\mathbf{u}_{kj}\|)$, where we have $\lbrace u_{kj}\rbrace_{j=1}^{m_k}$ knots in partition $k$. As in \eqref{globaleta} we can also provide a global prediction: 
\begin{equation}
\bm{\eta(\mathbf{s}^{\ast})}  =\sum_{k=1}^{K}[\mathbf{X}(\mathbf{s}^{\ast})^{\top}\bm{\beta}_k\mathbf{1}_{\{\mathbf{s}^{\ast}\in \mathcal{S}_k\}} + c_{k}(\mathbf{s}^{\ast})\bm{\widetilde{\Phi}}^{\top}_k(\mathbf{s}^{\ast})\bm{\delta}_k].
\label{globaletapred}    
\end{equation}
For given posterior samples $\lbrace \bm{\beta}_k, \bm{\delta}_k \rbrace_{k=1}^{K}$, we can obtain a posterior predictive distribution of $\bm{\eta(\mathbf{s}^{\ast})}$.

\subsection{Implementation Details}\label{SubSec:tune}

In this section, we provide automated heuristics for the tuning parameters. To implement SMB-SGLMMs, we need to specify the following components: (1) $K$ number of partitions, (2) location of knots in each partition, and (3) a weighting radius $\gamma$ for smoothing the local processes. In practice, we can set $K \leq C$ (number of available cores) for parallel computation. Our method is heavily parallelizable, so computational walltimes tend to decrease with larger $K$. 
However, selecting a very large $K$ may result in unreliable local estimates due to a small number of observations $N_k$ within each partition. In our simulation study, we compare the performance of our approach with varying $K$. Then, we select the $K$ that minimizes the out-of-sample root cross validated mean squared prediction error ($\mbox{rCVMSPE}$). Based on simulation results, the SMB-SGLMM is robust to the choice of $K$.  

To avoid overfitting, we use lasso \citep{tibshirani1996regression} to select the appropriate number and location of the knots. Initially, we set $m$ candidate knots $\lbrace \mathbf{u}_{kj} \rbrace_{j=1}^{m}$ uniformly over each partition $\mathcal{S}_k$ (e.g. $m \approx 1000$). Then we fit a penalized glm with lasso using response $\mathbf{Z}_k$ and covariates $[\mathbf{X}_k,  \mathbf{\Phi}_k]$, where $\mathbf{\Phi}_k$ is an $N_k$ by $m$ matrix. We impose an $l_1$ penalty to only the basis coefficients $\bm{\delta}_k$, not the fixed effects $\bm{\beta}_k$. We use the {\tt glmnet} package \citep{friedman2010regularization} in {\tt R} for lasso regression. For basis selection, we choose the basis functions corresponding to the nonzero basis coefficients. Since we run lasso regression independently for each partition, this step is embarrassingly parallel. 




From a pre-specified set of values (e.g., $\gamma=0.01,0.025,0.05,0.1$), we choose the $\gamma$ that yields the lowest $\mbox{rCVMSPE}$. Note that we choose $\gamma$ upon completion of Steps 1-3, the computationally demanding parts of SMB-SGLMM. Since the calculations in \eqref{globaletapred} are inexpensive, there are very little additional costs associated with Step 4.

\subsection{Computational Complexity}\label{SubSec:complexity}

We examine the computational complexity of SMB-SGLMM and illustrate how our approach scales with an increasing number of observations $N$. The three computationally demanding components are (1) basis selection (lasso), (2) MCMC for fitting the local processes, and (3) obtaining the global process. Here, parallelized computing is integral to the scalability of SMB-SGLMM. We provide the following discussion on computational costs and parallelization for each step:

\begin{enumerate}
    \item \textbf{Basis Selection:} In each partition, our methods select the $m_k$ knots from $m$ candidates using a regularization method (lasso). Based on results in \cite{friedman2010regularization} the cost of the coordinate descent-based lasso is $\mathcal{O}(N_k m)$, where $N_k$ is the number of observations in a partition $\mathcal{S}_k$. We can select the basis functions for each partition in parallel across $K$ processors. 
    \item \textbf{MCMC for local processes:} The computational cost is dominated by matrix-vector multiplications $\bm{\Phi}_k\bm{\delta}_k$, where $\bm{\Phi}_k$ is the $N_k$ by $m_k$ basis function matrix from the previous lasso step. The costs for this step is $\mathcal{O}(N_k m_k)$. We can fit the local processes in parallel across $K$ processors. 
    \item \textbf{Global Process:} We obtain the global process using weighted averages in \eqref{globalW}. This step requires $\mathcal{O}(N^2)$ complexity to calculate a distance matrix because the weights $c_k(\mathbf{s})$ in \eqref{globalW} are based on the distances between observations. Computing $c_k(\mathbf{s})$ requires a one-time computation of the distance matrix for all $N$ locations, which can be readily parallelized across $C$ available processors. We propose a novel way to ``stream'' the distances (Supplement) so that we can compute the weights $c_k(\mathbf{s})$ without actually storing the final distance matrix (e.g. $8\mbox{TB}$ for $N=1$ million). 
\end{enumerate}

Table~\ref{Complexity} summarizes complexity of SMB-SGLMM. Considering that the complexity of the stationary SGLMM is $\mathcal{O}(N^3)$, SMB-SGLMM is fast and provides accurate predictions for nonstationary processes (details in Sections~\ref{Sec:Simulation},\ref{Sec:Applications}).

\begin{table}[tt]
\centering
\begin{tabular}{ccc}
  \hline
 Operations & Complexity \\
  \hline
Basis selection & $\mathcal{O}(\sum_{k=1}^{K}N_k m/K)$  \\
MCMC & $\mathcal{O}(\sum_{k=1}^{K}N_k m_k/K)$ \\
Weighted average & $\mathcal{O}(N^2/C)$\\
   \hline
\end{tabular}
\caption{Computational complexity of SMB-SGLMMs. $K$ is the number of partitions and $C$ is the total available cores. $N_k$ is the number of observations, and $m_k$ is the number of knots from each partition. Knots are selected from $m$ candidate knots using a lasso.}
\label{Complexity} 
\end{table}

\section{Simulated Data Examples}\label{Sec:Simulation}

We implement SMB-SGLMMs in two simulated examples of massive ($N=100,000$) nonstationary binary and count data. We implement our approach in {\tt{nimble}} \citep{nimble2017}, a programming language for constructing and fitting Bayesian hierarchical models. Parallel computation is implemented through the {\tt{parallel}} package in {\tt{R}}.  The computation times are based on a single 2.2 GHz Intel Xeon E5-2650v4 processor. All the code was run on the Pennsylvania State University Institute for Cyber Science-Advanced Cyber Infrastructure (ICS-ACI) high-performance computing infrastructure. Source code is provided in the Supplement. 

Data is generated on $125,000$ locations on the spatial domain $\mathcal{S}\in\mathbb{R}^{2}$. We fit the spatial models using $N=100,000$ observations and reserve the remaining $N_{cv}=25,000$ observations for validation. We denote the model-fitting observations as $\mathbf{Z}=\{Z(s_{i}):s_{i}\in \mathbf{s}\}$ where $\mathbf{s}=\{s_{1},...,s_{N}\}$. Observations are generated using the SGLMM framework described in \eqref{Model} with $\bm{\beta}=(1,1)$. The nonstationary spatial random effects $\mathbf{W}=\{W(s_{i}):s_{i}\in \mathbf{s}\}$ are generated through convolving spatially varying kernel functions \citep{higdon1998process, paciorek2006spatial, risser2015local}. For some $s\in\mathbf{s}$ and reference locations $u_{j}\in \mathcal{D}$, we have $\mathbf{W}(s)=\sum_{j=1}^{J}K_{s}(u_{j})V(u_{j})$, where $K_{s}(u_{j})$ is a spatially varying Gaussian kernel function centered at reference location $u_{j}$ and $V(u_{j})$ is a realization of Gaussian white noise. Additional details are provided in the Supplement. The binary dataset uses a Bernoulli data model and a logit link function $\mbox{logit}(p)=\frac{p}{1-p}$, and the count dataset is similarly generated using a Poisson data model and a log link function.

We model the localized processes using the hierarchical framework in \eqref{LocationHierarchy}. To complete the hierarchical model, we set priors following  \citet{hughes2013dimension}: $\beta\sim N(\mathbf{0},100I)$ and $\sigma^2\sim IG(0.5,2000)$. We study SMB-SGLMM for different combinations of $K$ (the number of partitions) and $\gamma$ (weighting radius). We examine five partition groups $K=\{4, 9 , 16, 25, 36\}$ and four weighting radii $\gamma=\{0.1, 0.25, 0.5, 1\}$. In total, we study a total of $5 \times 4  = 20$ implementation. 

For each case, we perform basis selection via lasso using the {\tt{glmnet}} R package \citep{glmnet2010}. We generate $100,000$ samples from the posterior distribution $\pi(\bm{\beta},\bm{\delta},\sigma^2)$ using a block random-walk Metropolis-Hastings algorithm using the adaptation routine from \citet{shaby2010exploring}. We examine predictive ability and computational cost. These include $\mbox{rCVMSPE}=\sqrt{\frac{1}{N_{cv}}\sum_{i=1}^{N_{cv}}(Z_{i}-\hat{Z}_{i})^{2}}$ and  the walltime required to run  $100,000$ iterations of the MCMC algorithm. In addition, we present the posterior predictive intensity and probability surfaces.

\subsection{Count Data}\label{SubSec:SimulatedCount}

\begin{figure}
\begin{center}
\includegraphics[ scale = 0.4]{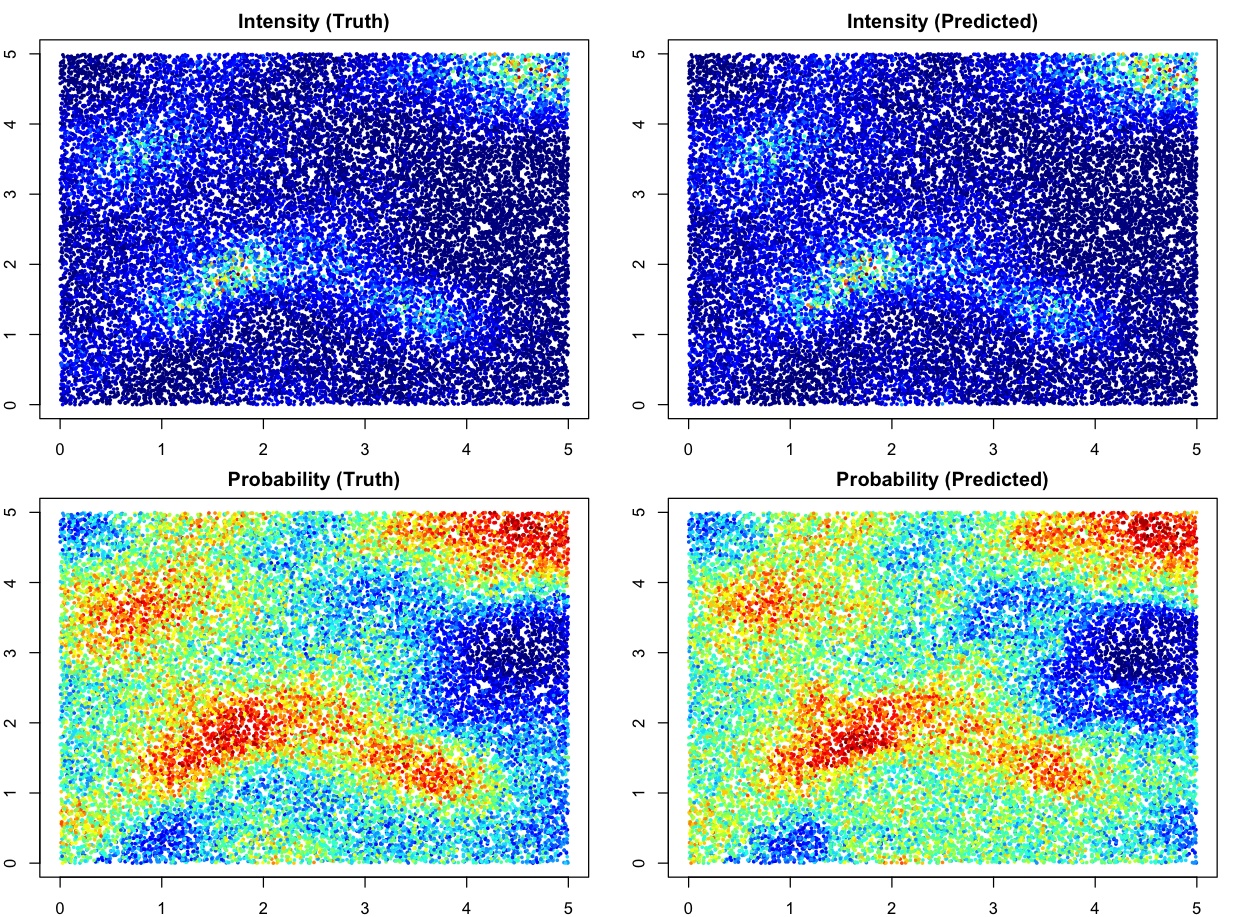}
\end{center}
\caption[]{True (top left) and predicted intensity surfaces (top right) for the simulated count data example. True (bottom left) and predicted probability surfaces (top right) for the simulated binary data example. We set $K=36$ and $\gamma=0.1$ for the count example and $K=25$ and $\gamma=0.1$ for the binary case.}
\label{ProbIntensity}
\end{figure}

Table~\ref{PoissonResults} presents results for the out-of-sample prediction errors for the SMB-SGLMM approach. Results indicate that the performance of our approach is robust across different combinations of $K$ and $\gamma$. For this example, predictive accuracy improves as we increase the number of partitions ($K$) and decrease the width of the weight radius ($\gamma=0.1$). We provide the posterior predictive intensity surface in Figure \ref{ProbIntensity} for the implementation yielding the lowest rCVMSPE ($K=36$ and $\gamma=0.1$). Based on visual inspection, the SMB-SGLMM approach captures the nonstationary behavior of the true latent spatial process. 

We report the combined walltimes for running lasso, MCMC, and weighting. Walltimes decrease considerably as we increase the number of partitions. This is not surprising as we fit these models in parallel and the sample size for each partition tends to decrease as we increase the number of total partitions. For each partition, model fitting incurs a computational cost of $\mathcal{O}(N_{k}m_{k})$ where $N_{k}$ and $m_{k}$ are the number of observations and the number of selected basis functions (thin-plate splines) within partition $k$, respectively. For the case where $K=36$, the median number of basis functions per partition is $14$ with a range of $4$ to $169$.

The localized parameter estimates of $\bm{\beta}$ are centered around the true parameter values $\bm{\beta}=(1,1)$ (Supplement). For the case where the number of partitions $K=36$, we also provide a map of the localized estimates of $\bm{\beta}=(\beta_{1},\beta_{2})$ in the Supplement. 

\begin{table}[tt]

\centering
\begin{tabular}{rrrrrr}
  \hline
   &  \multicolumn{4}{c}{Weighting Radius ($\gamma$)} & Walltime\\
 Partitions& 0.1 & 0.25 & 0.5 & 1 & (minutes)\\ 
  \hline
4 & 1.079 & 1.109 & 1.147 & 1.204 & 113.13 \\ 
  9 & 1.060 & 1.084 & 1.137 & 1.228 & 67.92 \\ 
  16 & 1.059 & 1.073 & 1.104 & 1.196 & 65.19 \\ 
  25 & 1.057 & 1.073 & 1.127 & 1.576 & 65.34 \\ 
  36 & \textbf{1.054} & 1.069 & 1.162 & 2.145 & 25.32 \\ 
   \hline
\end{tabular}
\caption{Cross-validation root mean squared prediction error (rCVMSPE) and total walltime (minutes) for the count data simulated example. Rows denote the five partition classes and columns correspond to the chosen weighting radius ($\gamma$). We report the combined walltime for lasso, MCMC, and weighting.}
\label{PoissonResults}
\end{table}

\subsection{Binary Data}\label{SubSec:SimulatedBinary}

In Table~\ref{BinaryResults}, we present prediction results for the binary simulated dataset. For this example, we observe that increasing the number of partitions $K$ and reducing the neighbor radius $\gamma$ results in more accurate predictions and lower computational costs. Figure \ref{ProbIntensity} includes the posterior predictive probability surface for the implementation yielding the lowest root CVMSPE ($K=25$ and $r=0.1$). For the case where $K=25$, the mean number of basis functions per partition is $7.5$ with a range of $0$ to $33$. In addition, computational walltimes decrease considerably as we increase the number of partitions for the reasons presented in Section \ref{SubSec:SimulatedCount}. The localized parameter estimates of $\bm{\beta}$ are centered around the true parameter values $\bm{\beta}=(1,1)$ (Supplement). For the case where the number of partitions $K=25$, we provide a map of the localized estimates of $\bm{\beta}=(\beta_{1},\beta_{2})$. 

\begin{table}[tt]
\centering
\begin{tabular}{rrrrrr}
  \hline
   &  \multicolumn{4}{c}{Weighting Radius ($\gamma$)} & Walltime\\
 Partitions& 0.1 & 0.25 & 0.5 & 1 & (minutes)\\ 
  \hline
4 & 0.360 & 0.362 & 0.371 & 0.380 &  70.15 \\ 
  9 & \textbf{0.348} & 0.351 & 0.365 & 0.381 &  17.76 \\ 
  16 & \textbf{0.348} & 0.351 & 0.370 & 0.402 &  15.98 \\ 
  25 & \textbf{0.348} & 0.350 & 0.365 & 0.397 &  8.65 \\ 
  36 & 0.349 & 0.350 & 0.364 & 0.402 &  16.330 \\ 
   \hline
\end{tabular}
\caption{Cross-validation root mean squared prediction error (rCVMSPE) and total walltime (minutes) for the binary simulated example. Rows denote the five partition classes and columns correspond to the chosen weighting radius ($\gamma$). We report the combined walltime for lasso, MCMC, and weighting.}
\label{BinaryResults}
\end{table}

\section{Applications} \label{Sec:Applications}

In this section, we apply our method to two real-world datasets pertaining to malaria incidence in the African Great Lakes region and cloud cover from satellite imagery. For both large non-Gaussian nonstationary datasets, SMB-SGLMM provides accurate predictions within a reasonable timeframe.

\subsection{Malaria Incidence in the African Great Lakes Region}\label{SubSec:malaria}

Malaria is a parasitic disease which can lead to severe illnesses and even death. Predicting occurrences at unknown locations can be of significant interest for effective control interventions. We compiled malaria incidence data from the Demographic and Health Surveys of 2015 \citep{ICF2017}. The dataset contains malaria incidence (counts) from $4,741$ GPS clusters in nine contiguous countries in the African Great Lakes region: Burundi, the Democratic Republic of Congo, Malawi, Mozambique, Rwanda, Tanzania, Uganda, Zambia, and Zimbabwe. We use the population size, average annual rainfall, vegetation index of the region, and the proximity to water as spatial covariates. Under a spatial regression framework, \cite{gopal2019characterizing} analyzes malaria incidence in Kenya using these environmental variables. In this study, we extend this approach to multiple countries in the African Great Lakes region. 

We use $N=3,973$ observations to fit the model and save $N_{cv}=948$ observations for cross-validation. We study the performance of SMB-SGLMMs for different combinations of $K\in \{2, 3, 4\}$ and $\gamma \in \{0.035,0.075,0.1,0.2,0.3\}$. For each partition, we set the number of candidate knots to be approximately $m=500$ and perform basis selection using lasso \citep{tibshirani1996regression}. On average $49.5$ basis functions are selected per partition. We fit a local spatial model \eqref{LocationHierarchy} running the MCMC algorithm for $200,000$ iterations. 

\begin{table}[tt]
\centering
\begin{tabular}{rrrrrrr}
  \hline
     &  \multicolumn{5}{c}{Weighting Radius ($\gamma$)} & Walltime\\
 Partitions & 0.035 & 0.075 & 0.1 & 0.2 & 0.3 & (minutes) \\ 
  \hline
2 & 55.1 & 62.7 & 69.4 & 78.6 & 88.2 & 33.2 \\ 
  3 & 58.6 & 69.7 & 76.8 & 95.2 & 132.1 & 19.9 \\ 
  4 & 56.9 & 70.7 & 82.1 & 82.2 & 94.2 & 16.9 \\ 
   \hline
\end{tabular}
\caption{Root CVMSPE and total walltime (mins) for the malaria incidence example. Rows denote the four partition classes and columns correspond to the chosen weighting radius. We report the combined walltime for lasso, MCMC, and weighting.} 
\label{Table:malaria}
\end{table}


\begin{figure}
\begin{center}
\includegraphics[ scale = 0.5]{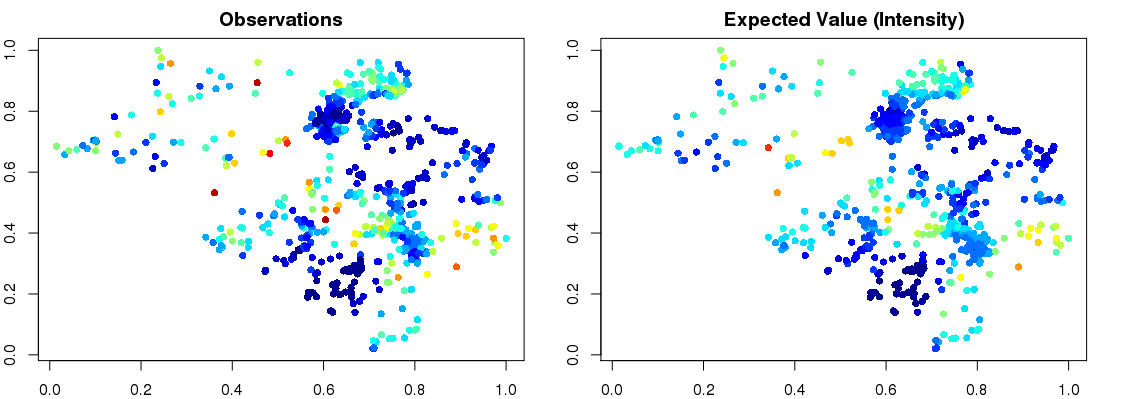}
\end{center}
\caption[]{Illustration of the malaria occurrence dataset for $K=2$ and $\gamma=0.035$. True observations (left) and posterior predictive intensity surface (right) for the validation sample.}
\label{MalariaIntensity}
\end{figure}

\begin{figure}
\begin{center}
\includegraphics[ scale = 0.7]{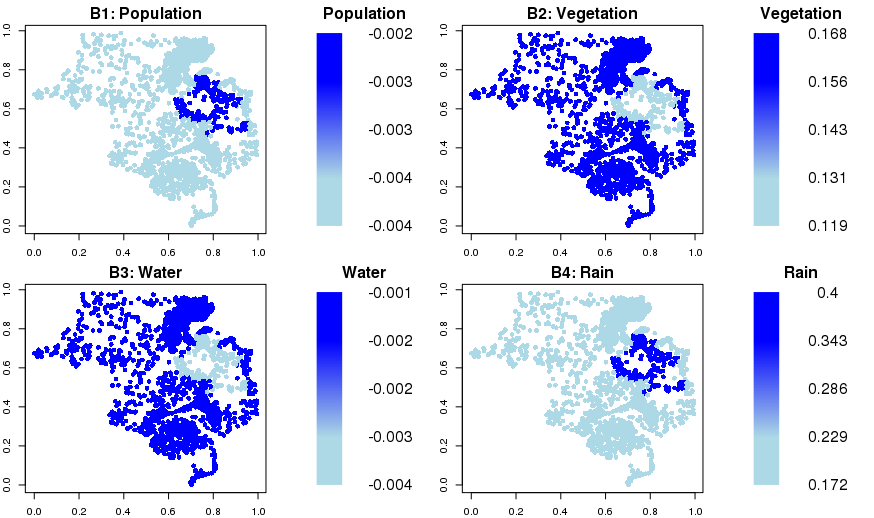}
\end{center}
\caption[]{Posterior mean estimates of partition-varying ($K=2$) fixed effects $\bm{\hat{\beta}}$. Estimated coefficients $(\{\hat{\beta}_{1}, \hat{\beta}_{2}, \hat{\beta}_{3}, \hat{\beta}_{4}\})$ for covariates population (top left), vegetation (top right), water (bottom left), and rain (bottom right).}
\label{MalariaBeta}
\end{figure}

Table~\ref{Table:malaria} compares the rCVMSPE for each case. We observe that rCVMSPE increases with larger weighting radii $\gamma$, possibly due to over smoothing in the partition boundaries. For smaller $\gamma$ ($0.035$ and $0.075$), we find that predictions are not sensitive to the choice of $K$. For this example, setting $K=2, \gamma=0.035$ yields the most accurate predictions. In Figure~\ref{MalariaIntensity}, the predicted intensities of the validation locations exhibit similar spatial patterns as the true count observations. We provide maps for the partition-varying coefficients ($K=2$) in Figure~\ref{MalariaBeta}. The smaller partition includes parts of northern Malawi, southern Tanzania, and northeastern Zambia. Here, the values of $\widehat{\beta}>0$ indicate that the corresponding covariates have a positive relationship with malaria incidence. From the estimates of $\widehat{\beta}_4$, we observe that while rainfall may increase malaria incidence, these effects are more pronounced in the smaller partition.

\subsection{Moderate Resolution Imaging Spectroradiometer (MODIS) Cloud Mask Data}\label{SubSec:MODIS}

The National Aeronautics and Space Administration (NASA) launched the Terra Satellite in December 1999 as part of the Earth Observing System. As in past studies \citep{sengupta2013hierarchical, bradley2019bayesian}, we model the cloud mask data captured by the Moderate Resolution Imaging Spectroradiometer (MODIS) instrument onboard the Terra satellite. The response is a binary incidence of cloud mask at a 1km $\times$ 1km spatial resolution. In this study, we selected $N=1,000,000$ observations to fit our model and reserved $N_{cv}=111,000$ for validation. We model the binary observations as a nonstationary SGLMM via the SMB-SGLMM method. Similar to \citet{sengupta2013hierarchical, bradley2019bayesian}, we include the vector $\mathbf{1}$ and a vector of latitudes as the covariates and use a logit link function. 

For the SMB-SGLMM approach, we vary the number of partitions $K\in \{16, 25, 36, 49\}$ and weighting radius $\gamma \in \{0.01,0.025,0.05,0.1\}$ for a total of $16$ cases. For each partition, we begin with $m=1,000$ knots (candidates) and perform basis selection using lasso regression \cite{tibshirani1996regression}. On average, basis selection results in roughly $16.3$ basis functions per partition. For each partition, we fit a localized spatial basis SGLMM \eqref{LocationHierarchy} by running the MCMC algorithm for $100,000$ iterations. 

\begin{figure}
\begin{center}
\includegraphics[ scale = 0.7]{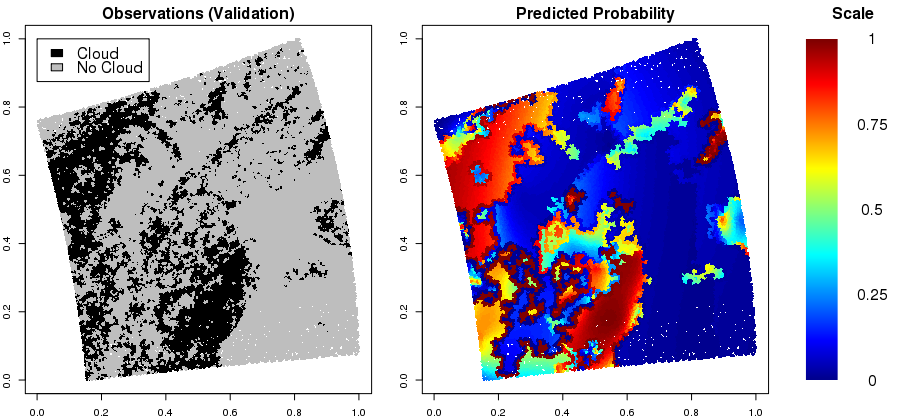}
\end{center}
\caption[]{Illustration for the MODIS cloud mask dataset. (left) True observations of cloud mask. (right) Posterior predictive probability surface ($K=25$ and $\gamma=0.01$).}
\label{MODISprobability}
\end{figure}

Figure~\ref{MODISprobability} indicates that there are similar spatial patterns between binary observations and predicted probability surface. We also provide the misclassification rate for each case in Table~\ref{Table:MODIS}. The performance of SMB-SGLMMs is robust across different choices of $K$ and $\gamma$. Results suggest that a moderate number of partitions ($K=25$) and a smaller weighting radius ($\gamma=0.01$) yields the most accurate predictions. The combined walltimes decrease when using more partitions; however, these are on the order of hours in all cases.

\begin{table}[tt]
\centering
\begin{tabular}{rrrrrrrrrrr}
  \hline
  &  \multicolumn{4}{c}{Weighting Radius ($\gamma$)} & Walltime\\
 Partitions &  0.01 & 0.025 & 0.05 & 0.1 & (hours) \\ 
  \hline
16 & 0.192 & 0.215 & 0.229 & 0.266 & 5.4 \\ 
  25 & \textbf{0.174} & 0.175 & 0.184 & 0.211 & 3.9 \\ 
  36 & 0.182 & 0.200 & 0.223 & 0.299 & 3.2 \\ 
  49 & 0.186 & 0.198 & 0.222 & 0.290 & 2.8 \\
  
 \hline  
\end{tabular}
\caption{Misclassification Rate and total walltime (hours) for the MODIS cloud mask example. Rows denote the four partition classes and columns correspond to the chosen weighting radius. We report the combined walltime for lasso, MCMC, and weighting.} 
\label{Table:MODIS}
\end{table}

\section{Discussion}\label{Sec:Discussion}
In this manuscript, we propose a scalable algorithm for modeling massive nonstationary non-Gaussian datasets. Existing approaches are limited to either stationary non-Gaussian or nonstationary Gaussian spatial data, but not both. Our method divides the spatial domain into disjoint partitions using a spatial clustering algorithm \citep{heaton2017nonstationary}. For each partition, we fit a localized model using a collection of thin plate spline basis functions. Here, the linear combinations of the basis functions capture the underlying nonstationary behavior. We provide an automated basis selection process via a regularization approach, such as lasso. This framework is computationally efficient due to parallel computing and using basis representations of complex spatial processes. Our study shows that the proposed method provides accurate estimations and predictions within a reasonable time. Moreover, our approach scales well to massive datasets, where we model $N=1$ million binary observations within 4 hours. To our knowledge, this is the first method geared towards modeling nonstationary non-Gaussian spatial data at this scale. 



The proposed framework can be extended to a wider range of spatial basis functions. In the literature, there exists a wide array of spatial basis functions such as bi-square (radial) basis functions using varying resolutions \citep{cressie2008fixed,nychka2015multiresolution,katzfuss2017multi}, empirical orthogonal functions \citep{cressie2015statistics}, predictive processes \citep{banerjee2008gaussian}, Moran's basis functions \citep{griffith2003spatial, hughes2013dimension}, wavelets  \citep{nychka2002multiresolution,shi2007global}, Fourier basis functions \citep{royle2005efficient} and Gaussian kernels \citep{higdon1998process}. A closer examination of adopting Bayesian regularization methods (see \cite{o2009review} for a detailed review) for selecting basis functions is also an interesting future research avenue. 

Developing scalable methods for modeling nonstationary non-Gaussian spatio-temporal data is challenging. The partition-based basis function representation can be integrated into existing hierarchical spatio-temporal models. For example, we can approximate the nonstationary processes using a tensor product of spatial and temporal basis functions or by constructing data-driven space-time basis functions.

\section*{Acknowledgements}\label{Sec:Acknowledgements}
Jaewoo Park was supported by the Yonsei University Research Fund 2020-22-0501 and the National Research Foundation of Korea (NRF-2020R1C1C1A0100386811). The authors are grateful to Matthew Heaton, Murali Haran, John Hughes, and Whitney Huang for providing useful sample code and advice. The authors are thank the anonymous reviewers for their careful review and valuable comments.

\clearpage
\appendix

\title{\textbf{Supplementary Material for A Scalable Partitioned Approach to Model Massive Nonstationary Non-Gaussian Spatial Datasets}}

\section{Spatial Clustering Algorithm}

Here, we provide the clustering algorithm \citep{heaton2017nonstationary} in detail. We obtain residuals $\bm{\epsilon}$ from a GLM fit with a response vector $\mathbf{Z}\in \mbR^{N}$ and a covariate matrix $\mathbf{X} \in \mbR^{N\times p}$. Let $\bm{\epsilon}_{k} \in \mbR^{N_{k}}$ be the residuals belongs to the cluster (partition) $\mathcal{S}_k$. Then we can define the dissimilarity between two clusters as 
\[
d(\mathcal{S}_{k_1},\mathcal{S}_{k_2})=\Big[\frac{N_{k_1}N_{k_2}}{N_{k_1}+N_{k_2}}(\bar{\epsilon}_{k_1}-\bar{\epsilon}_{k_2})^2\Big]\frac{1}{\bar{E}},
\]
where $\bar{\epsilon}_k$ is the average of $\bm{\epsilon}_{k}$ and $\bar{E}$ is the average Euclidean distance between points in $\mathcal{S}_{k_1}, \mathcal{S}_{k_2}$. Then the spatial clustering algorithm can be summarized as follows. 

\begin{algorithm}
\caption{Spatial clustering algorithm \citep{heaton2017nonstationary} }\label{clusteringalg}
\begin{algorithmic}[H]
\normalsize
\State Initialize each location $\mathbf{s}_k=\mathcal{S}_k$ for $k=1,\cdots,N$; we have $N$ number of clusters.

\State 1. Find clusters $\mathcal{S}_{k_1}$,$\mathcal{S}_{k_1}$ having the minimum
$d(\mathcal{S}_{k_1},\mathcal{S}_{k_2})$ where $\mathbf{s}_i \sim \mathbf{s}_j$ (Voronori neighbors) for $\mathbf{s}_i \in \mathcal{S}_{k_1}$ and $\mathbf{s}_j \in \mathcal{S}_{k_2}$

\State 2. Combine two clusters $$\mathcal{S}_{\min\lbrace k_1,k_2 \rbrace} = \mathcal{S}_{k_1} \cup \mathcal{S}_{k_2}$$ and set $$\mathcal{S}_{\max\lbrace k_1,k_2 \rbrace} =\emptyset$$

\State Repeat 1-2 until we have $K<N$ number of clusters.
\end{algorithmic}
\end{algorithm}

We note that Algorithm~\ref{clusteringalg} becomes computationally expensive with increasing number of observations. Following suggestions in \cite{heaton2017nonstationary}, we perform clustering after combining observations to a lattice $\lbrace \mathbf{s}^{\ast}_{l} \rbrace_{l=1}^{L}$ ($L<<N$). Here, $\mathcal{N}_{l}=\lbrace \mathbf{s}_i: \|\mathbf{s}_i-\mathbf{s}^{\ast}_{l}\| < \|\mathbf{s}_i-\mathbf{s}^{\ast}_{m}\|\rbrace$ is the collection of observations whose closest lattice point is $\mathbf{s}^{\ast}_{l}$, and $\bar{\epsilon}(\mathbf{s}^{\ast}_{l})= |\mathcal{N}_l|^{-1}\sum_{s_i \in \mathcal{N}_{l}} \epsilon(\mathbf{s}_{i})$. Then we apply Algoritm~\ref{clusteringalg} to $\lbrace \bar{\epsilon}(\mathbf{s}^{\ast}_{l}) \rbrace_{l=1}^{L}$ rather than to $\lbrace \epsilon(\mathbf{s}_{i}) \rbrace_{i=1}^{N}$. Since the number of lattice points $L$ is much smaller than the number of observations $N$, spatial clustering algorithm becomes computationally feasible. For instance, in our simulation studies we chose $L=900$ for $N=100,000$.

\section{Simulation of Nonstationary Spatial Random Effects}
We describe how to generate the nonstationary spatial random effects from Section 4,  $\mathbf{W}=\{W(s_{1}),...,W(s_{n}))\}$. The nonstationary spatial random effects $\mathbf{W}$ are generated by convolving a collection of spatially varying kernel functions \citep{higdon1998process, paciorek2006spatial, risser2015local}. The construction procedure is broken down into four steps: (1) select locations for the ``basis'' and ``reference'' kernel functions; (2) construct ``basis'' kernels; (3) use ``basis'' kernels to construct ``reference'' kernels on a finite grid; and (4) generate non-stationary spatial random effects using the ``reference kernels''. 

\begin{figure}
\begin{center}
\includegraphics[ scale = 0.5]{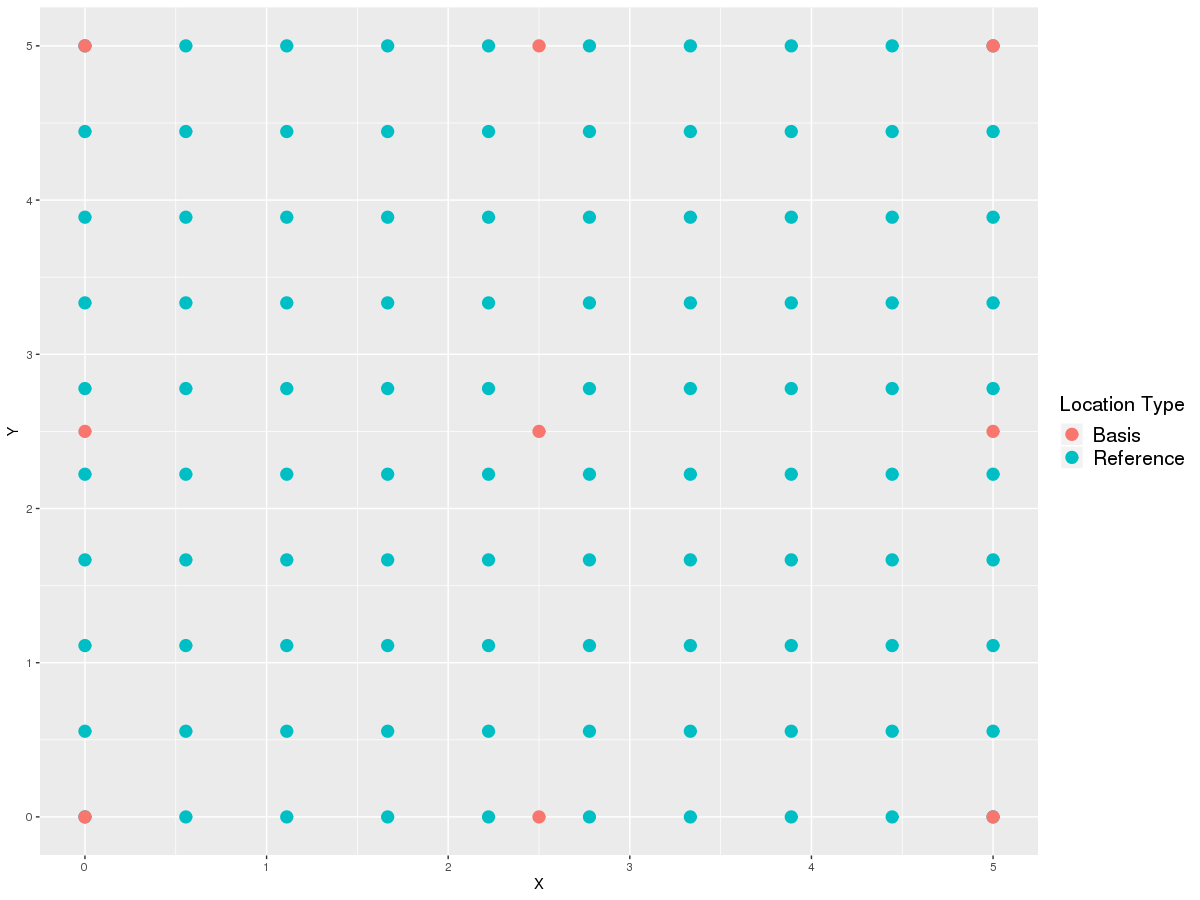}
\end{center}
\caption[]{Locations (knots) for ``basis'' (red) and ``reference'' (blue) locations.  }
\label{SFigure:Locations}
\end{figure}

In Step 1, we select $M$ ``basis'' locations $\mathbf{b}=\{b_1,...,b_M\}$ on a coarse grid of evenly-spaced locations over the spatial domain $\mathcal{D}\in\mathbb{R}^{2}$. Similarly, we select $J$ ``reference'' locations $\mathbf{u}=\{u_1,...,u_J\}$ on a finer grid of evenly-space locations in $\mathcal{D}$. As in past studies \citep{higdon1998process, paciorek2006spatial, risser2015local}, we typically select $M<J$. Figure \ref{SFigure:Locations} illustrates the placement of the `basis'' and ``reference'' locations. 

In Step 2, we construct Gaussian ``basis'' kernels centered at each basis location $\mathbf{b}$ for $m=1,...,M$. The ``basis'' kernels are defined as 
$$K_{m}(x_{i})=(2\pi)^{-1}|\Sigma_{m}|^{-1/2}\exp\big\{-\frac{1}{2}(x_{i}-b_{m})'\Sigma_{m}(x_{i}-b_{m})\big\},$$
where $b_{m}$ is a ``basis'' location, $x_{i}$ is the location of interest, and $\Sigma_{m}$ is a $2\times 2$ covariance matrix for the $m$-th Gaussian ``basis'' kernel. 

In Step 3, we construct the Gaussian kernels for the reference locations $K_{s}(u_{j})$ for $j=1,...,J$ as a weighted average of the ``basis'' kernels $K_{m}(\cdot)$. The ``reference'' kernels are defined as:
$$K_{s}(u_{j})=\sum_{m=1}^{M}w_{m}(s)K_{m}(u_{j}),$$
where $w_{m}(s)$ are the distance-based weights, $u_{j}$ are the ``reference'' locations, and $s$ is the spatial location of interest. Here, the weights $w_{m}(s)\propto \exp\big\{-\frac{1}{2}||s-b_{m}||\big\}$ and $\sum_{m=1}^{M}w_{m}(s)=1$.

Finally, in Step 4, we generate the nonstationary spatial random effects as $$W(s)=\sum_{j=1}^{J}K_{s}(u_{j})V(u_{j}),$$
where $K_{s}(u_{j})$ is a spatially varying Gaussian kernel function centered at ``reference'' locations $u_{j}$ and $V(u_{j})$ is a realization of Gaussian white noise. Note that $V(u_{j})\sim \mathcal{N}(0,\sigma^{2}_u)$.

In our implementation, we chose $M=9$ ``basis'' locations $\mathbf{b}=\{b_1,...,b_M\}$ and $J=100$ ``reference'' locations $\mathbf{u}=\{u_1,...,u_J\}$ on a grid of evenly-space locations in $\mathcal{D}$. The ``basis'' kernel functions $K_{m}(\cdot)$ for $m=1,...,9$ have spatially-varying covariance matrices $\Sigma_{m}$ as follows:

$$\Sigma_{1}=\begin{bmatrix}
0.50 & 0.30\\
0.30 &0.33
\end{bmatrix} \qquad 
\Sigma_{2}=\begin{bmatrix}
0.50 & -0.12\\
-0.12 & 0.13
\end{bmatrix}\qquad 
\Sigma_{3}=\begin{bmatrix}
0.50 & 0.18\\
0.18 & 0.20
\end{bmatrix}$$

$$\Sigma_{4}=\begin{bmatrix}
0.50 & 0.54\\
0.54 & 0.60
\end{bmatrix} \qquad 
\Sigma_{5}=\begin{bmatrix}
0.50 & 0.06\\
0.06 & 0.07
\end{bmatrix}\qquad 
\Sigma_{6}=\begin{bmatrix}
0.50 & -0.48\\
-0.48 & 0.53
\end{bmatrix}$$

$$\Sigma_{7}=\begin{bmatrix}
0.50 & 0.42\\
0.42 & 0.46
\end{bmatrix} \qquad 
\Sigma_{8}=\begin{bmatrix}
0.50 & -0.36\\
-0.36 & 0.40
\end{bmatrix}\qquad 
\Sigma_{9}=\begin{bmatrix}
0.50 & -0.24\\
-0.24 & 0.26
\end{bmatrix}$$

\begin{figure}
\begin{center}
\includegraphics[ scale = 0.4]{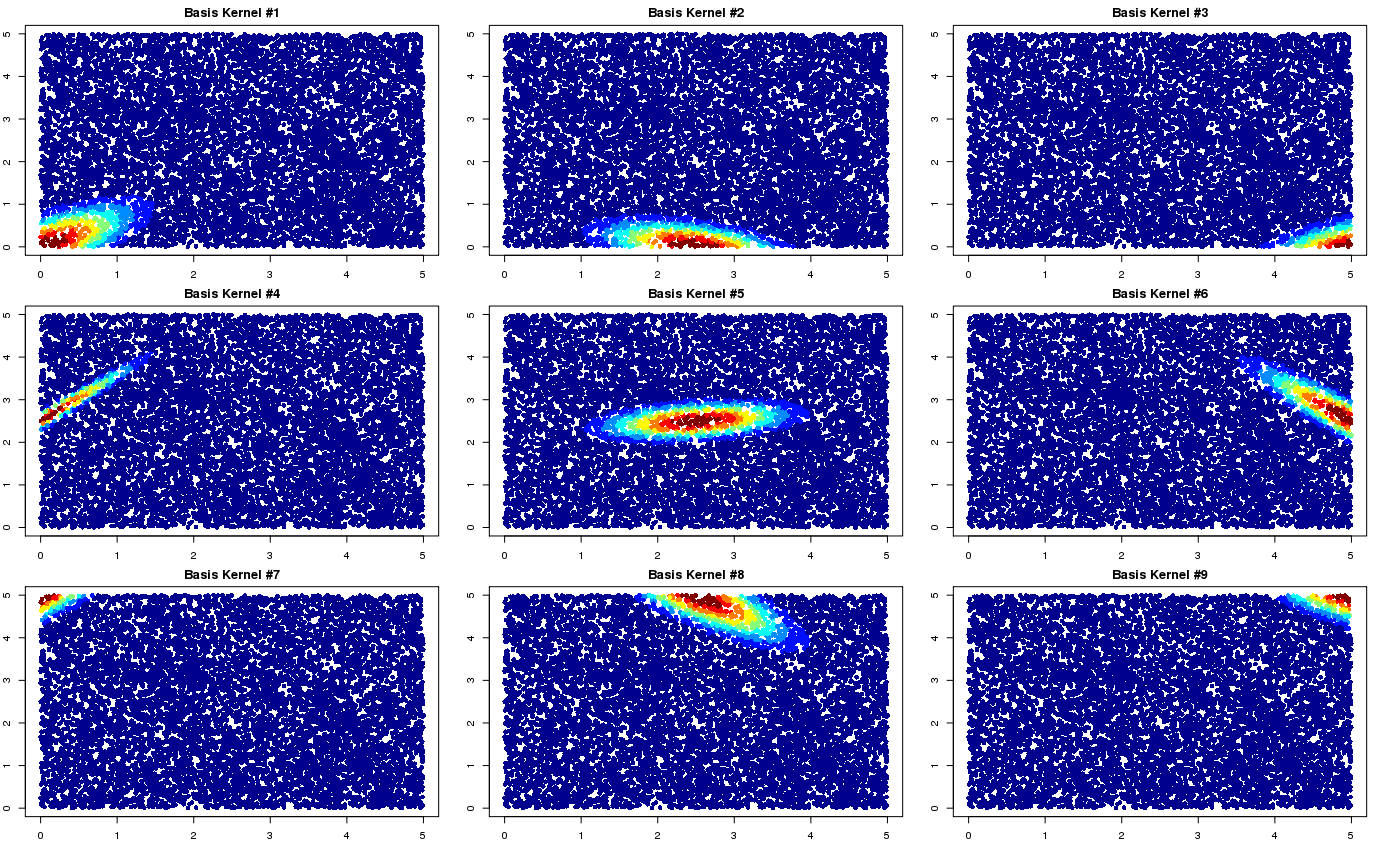}    
\end{center}
\caption[]{``Basis'' Kernel functions $K_{m}(\cdot)$ for locations $b_{m}$, $m=1,...,9$.}
\label{SFigure:BasisKernels}
\end{figure}

\begin{figure}
\begin{center}
\includegraphics[ scale = 0.4]{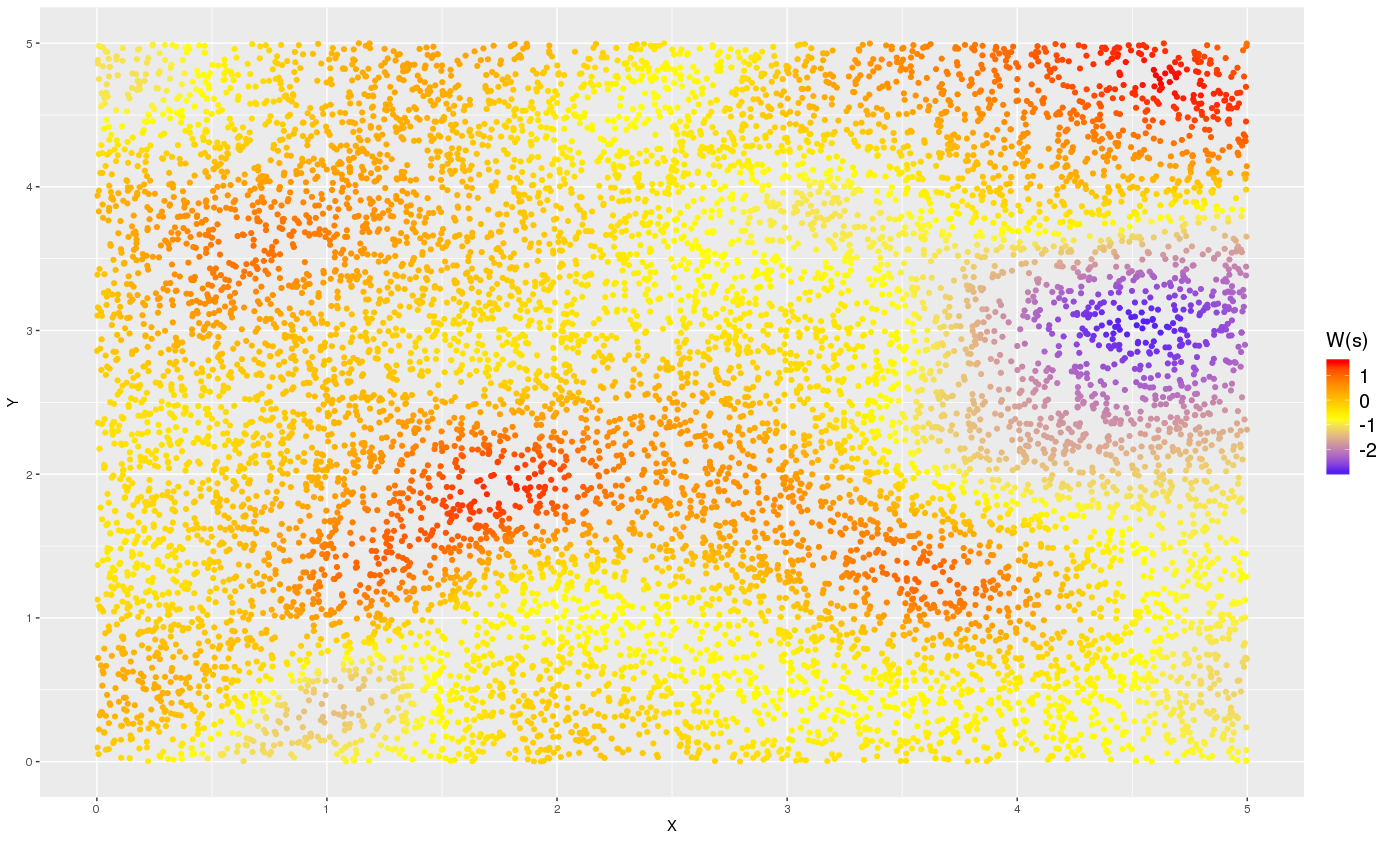}    
\end{center}
\caption[]{Nonstationary spatial random field constructed via convolutions of Gaussian basis kernels.}
\label{SFigure:Convolution}
\end{figure}

\section{Computing weights via parallelization}
As mentioned in Section 3.1 (step 4), we propose a parallelized method to compute weights $c_k(\mathbf{s})$ for $k=1,...,K$ and $\mathbf{s}\in \mathcal{S}$. For a given location $\mathbf{s}$, the weights $c_k(\mathbf{s}) \propto \exp\Big(-\|\mathbf{s}-\widetilde{\mathbf{s}}_{k}\|^2\Big)$ where $\widetilde{\mathbf{s}}_{k} \in \mathcal{S}_k$ is the point in partition $k$ with the shortest distance to point $\mathbf{s}$. The challenge lies in computing and storing the distances $\|\mathbf{s}-\widetilde{\mathbf{s}}_{k}\|$. Naive implementations may simply compute the distance matrix between all $N$ locations, which requires $\mathcal{O}(n^{2})$ operations as well as $\mathcal{O}(n^{2})$ in storage. In the MODIS example (Section 5.2), the distance matrix ($n=1$million observations) demands $8TB$ of storage. 

We propose ``streaming'' the weight calculations ($c_k(\mathbf{s})$) without storing the distance matrix. First, we parallelize over $C$ available cores where each core is assigned a location $\bm{s}$. Then, we compute distances between location $\bm{s}$ and the other locations $\bm{s}_{-}$. This n-dimensional vector of distances (e.g. $8MB$ for the MODIS case) can be stored in the random access memory (RAM). Finally, we compute the weights $c_k(\mathbf{s})$ for each partition $k$. The task concludes once $c_k(\mathbf{s})$ are computed for all $\bm{s}$ in our dataset and for all partitions $k$.

\section{Partition Varying Estimates}

\begin{figure}\label{PoissonBeta}
\begin{center}
\includegraphics[ scale = 0.7]{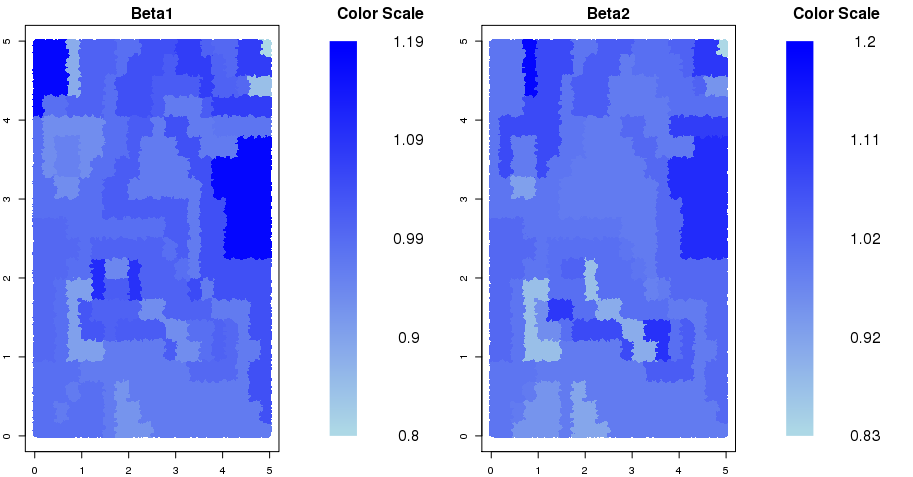}
\end{center}
\caption[]{Map of the spatially-varying $\beta_{1}$ parameter (left) and $\beta_{2}$ parameter (right) for the simulated Poisson Dataset)}
\end{figure}

\begin{figure}\label{BinaryBeta}
\begin{center}
\includegraphics[ scale = 0.7]{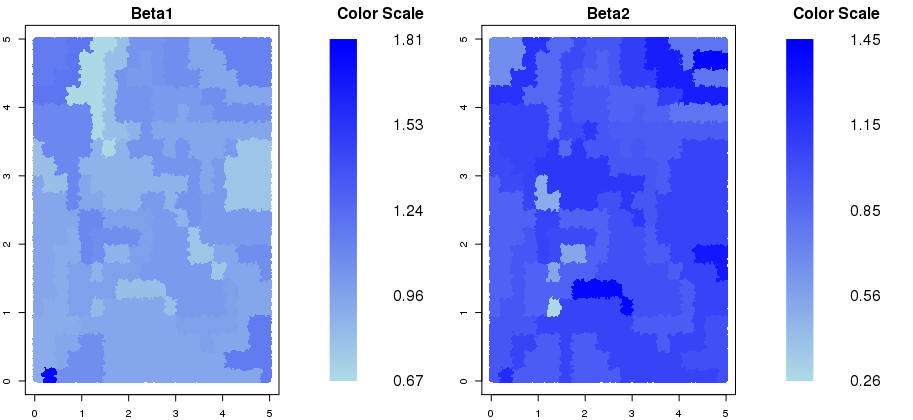}
\end{center}
\caption[]{Map of the spatially-varying $\beta_{1}$ parameter (left) and $\beta_{2}$ parameter (right) for the simulated Binary Dataset)}
\end{figure}

\begin{table}[ht]
\centering
\begin{tabular}{ccccc}
  \hline
  &  \multicolumn{2}{c}{Count Data} &  \multicolumn{2}{c}{Binary Data} \\
 Partition & $\beta_{1}$ & $\beta_{2}$ & $\beta_{1}$ & $\beta_{2}$ \\ 
  \hline
1 & 0.93 (0.87,0.98) & 0.98 (0.93,1.04) & 0.99 (0.96,1.02) & 1.02 (0.98,1.05) \\ 
  2 & 1.02 (0.92,1.13) & 0.98 (0.87,1.08) & 1.01 (0.98,1.06) & 1 (0.96,1.04) \\ 
  3 & 0.95 (0.72,1.18) & 1.08 (0.86,1.31) & 1.03 (0.96,1.1) & 1 (0.94,1.07) \\ 
  4 & 1.03 (0.81,1.24) & 1.15 (0.95,1.37) & 1 (0.93,1.08) & 1.05 (0.98,1.12) \\ 
   \hline
\end{tabular}
\caption{For $K=4$ partitions, we report parameter estimates for $\beta_{1}$ and $\beta_{2}$ for the simulated count and binary datasets. This includes the posterior mean and 95\% credible intervals for each partition.}
\end{table}

\begin{table}[ht]
\centering
\begin{tabular}{ccccc}
  \hline
  &  \multicolumn{2}{c}{Count Data} &  \multicolumn{2}{c}{Binary Data} \\
 Partition & $\beta_{1}$ & $\beta_{2}$ & $\beta_{1}$ & $\beta_{2}$ \\ 
  \hline
1 & 0.91 (0.77,1.06) & 0.97 (0.83,1.11) & 0.99 (0.94,1.03) & 1.01 (0.97,1.06) \\ 
  2 & 0.97 (0.88,1.05) & 1.03 (0.95,1.12) & 1 (0.95,1.05) & 1 (0.95,1.05) \\ 
  3 & 0.94 (0.64,1.25) & 0.96 (0.65,1.28) & 1.08 (0.97,1.2) & 1.05 (0.94,1.17) \\ 
  4 & 1 (0.7,1.31) & 0.95 (0.65,1.24) & 0.97 (0.87,1.08) & 0.97 (0.86,1.08) \\ 
  5 & 1.02 (0.92,1.12) & 0.98 (0.87,1.08) & 1.01 (0.95,1.07) & 1.03 (0.97,1.09) \\ 
  6 & 0.95 (0.72,1.17) & 1.08 (0.86,1.31) & 1.03 (0.96,1.1) & 1 (0.93,1.07) \\ 
  7 & 0.94 (0.84,1.03) & 1.02 (0.92,1.12) & 0.94 (0.86,1.02) & 1.05 (0.96,1.13) \\ 
  8 & 1.03 (0.82,1.24) & 1.15 (0.95,1.37) & 1.06 (0.98,1.15) & 1.02 (0.93,1.11) \\ 
  9 & 1.16 (0.79,1.54) & 0.69 (0.31,1.05) & 1 (0.93,1.07) & 1.05 (0.97,1.12) \\ 
   \hline
\end{tabular}
\caption{For $K=9$ partitions, we report parameter estimates for $\beta_{1}$ and $\beta_{2}$ for the simulated count and binary datasets. This includes the posterior mean and 95\% credible intervals for each partition.}
\end{table}
\begin{table}[ht]
\centering
\begin{tabular}{ccccc}
  \hline
  &  \multicolumn{2}{c}{Count Data} &  \multicolumn{2}{c}{Binary Data} \\
 Partition & $\beta_{1}$ & $\beta_{2}$ & $\beta_{1}$ & $\beta_{2}$ \\ 
  \hline
1 & 0.94 (0.72,1.15) & 1 (0.79,1.22) & 0.99 (0.93,1.04) & 1.03 (0.97,1.08) \\ 
  2 & 1.06 (0.7,1.41) & 0.95 (0.58,1.3) & 1 (0.95,1.05) & 0.99 (0.95,1.04) \\ 
  3 & 0.98 (0.89,1.07) & 1.02 (0.93,1.11) & 1.08 (0.97,1.19) & 1.06 (0.95,1.17) \\ 
  4 & 0.93 (0.62,1.24) & 0.98 (0.66,1.28) & 0.97 (0.86,1.08) & 0.97 (0.86,1.08) \\ 
  5 & 0.99 (0.69,1.29) & 0.95 (0.67,1.26) & 1 (0.88,1.12) & 1.03 (0.91,1.15) \\ 
  6 & 0.91 (0.72,1.11) & 0.94 (0.75,1.14) & 0.91 (0.78,1.04) & 0.87 (0.75,1) \\ 
  7 & 1.02 (0.92,1.13) & 0.98 (0.87,1.09) & 1 (0.85,1.15) & 1.12 (0.97,1.27) \\ 
  8 & 0.83 (0.44,1.21) & 1.44 (1.02,1.82) & 1.02 (0.95,1.08) & 1.02 (0.95,1.09) \\ 
  9 & 1.02 (0.66,1.35) & 0.91 (0.56,1.27) & 1.03 (0.95,1.11) & 1 (0.93,1.08) \\ 
  10 & 1.02 (0.75,1.29) & 1.08 (0.81,1.35) & 1.03 (0.88,1.18) & 1.02 (0.88,1.17) \\ 
  11 & 0.98 (0.86,1.1) & 1.03 (0.91,1.15) & 0.94 (0.86,1.03) & 1.05 (0.96,1.13) \\ 
  12 & 0.79 (0.4,1.21) & 1.1 (0.68,1.51) & 1.18 (0.97,1.39) & 1 (0.8,1.21) \\ 
  13 & 0.86 (0.69,1.03) & 1 (0.84,1.18) & 1.08 (0.97,1.19) & 1.02 (0.91,1.13) \\ 
  14 & 0.96 (0.69,1.23) & 1.29 (1.02,1.57) & 1.02 (0.87,1.16) & 1.02 (0.87,1.17) \\ 
  15 & 1.16 (0.78,1.54) & 0.69 (0.31,1.05) & 0.95 (0.85,1.06) & 1.02 (0.91,1.12) \\ 
  16 & 1.13 (0.79,1.47) & 0.96 (0.63,1.29) & 1.05 (0.96,1.15) & 1.09 (0.99,1.18) \\ 
   \hline
\end{tabular}
\caption{For $K=16$ partitions, we report parameter estimates for $\beta_{1}$ and $\beta_{2}$ for the simulated count and binary datasets. This includes the posterior mean and 95\% credible intervals for each partition.}
\end{table}

\begin{table}[ht]
\centering
\begin{tabular}{ccccc}
  \hline
  &  \multicolumn{2}{c}{Count Data} &  \multicolumn{2}{c}{Binary Data} \\
 Partition & $\beta_{1}$ & $\beta_{2}$ & $\beta_{1}$ & $\beta_{2}$ \\ 
  \hline
1 & 0.94 (0.72,1.15) & 1.01 (0.79,1.22) & 0.99 (0.94,1.05) & 1.03 (0.97,1.08) \\ 
  2 & 1.07 (0.72,1.43) & 0.95 (0.59,1.32) & 0.99 (0.74,1.23) & 0.95 (0.69,1.2) \\ 
  3 & 0.96 (0.83,1.08) & 1.08 (0.95,1.21) & 1 (0.96,1.05) & 1 (0.96,1.06) \\ 
  4 & 0.94 (0.63,1.25) & 0.96 (0.65,1.28) & 0.93 (0.75,1.12) & 0.92 (0.73,1.11) \\ 
  5 & 1.19 (0.85,1.52) & 1.03 (0.68,1.35) & 1.06 (0.93,1.19) & 1.03 (0.9,1.16) \\ 
  6 & 1 (0.7,1.3) & 0.95 (0.65,1.25) & 0.97 (0.86,1.08) & 0.97 (0.86,1.07) \\ 
  7 & 0.91 (0.71,1.1) & 0.94 (0.75,1.14) & 1 (0.88,1.11) & 1.03 (0.91,1.14) \\ 
  8 & 1.05 (0.91,1.19) & 0.98 (0.84,1.11) & 0.91 (0.78,1.03) & 0.87 (0.75,1) \\ 
  9 & 0.83 (0.44,1.21) & 1.44 (1.05,1.85) & 1 (0.85,1.14) & 1.12 (0.97,1.27) \\ 
  10 & 1.02 (0.8,1.25) & 1.03 (0.8,1.26) & 1.02 (0.94,1.08) & 1.02 (0.95,1.09) \\ 
  11 & 1.01 (0.73,1.3) & 1.02 (0.72,1.29) & 1.04 (0.88,1.2) & 0.96 (0.81,1.11) \\ 
  12 & 1.01 (0.67,1.36) & 0.92 (0.56,1.26) & 1.02 (0.85,1.18) & 1.11 (0.94,1.27) \\ 
  13 & 0.79 (0.36,1.23) & 0.96 (0.54,1.41) & 1.11 (0.91,1.31) & 1.01 (0.81,1.21) \\ 
  14 & 1.08 (0.55,1.59) & 1.33 (0.82,1.86) & 1.03 (0.88,1.18) & 1.02 (0.87,1.16) \\ 
  15 & 1.03 (0.75,1.29) & 1.08 (0.8,1.35) & 1.01 (0.89,1.14) & 0.96 (0.84,1.08) \\ 
  16 & 0.98 (0.86,1.11) & 1.03 (0.91,1.15) & 1.19 (0.95,1.42) & 1.14 (0.9,1.38) \\ 
  17 & 1.06 (0.8,1.32) & 0.97 (0.71,1.23) & 0.93 (0.7,1.16) & 0.93 (0.69,1.16) \\ 
  18 & 0.79 (0.37,1.19) & 1.09 (0.68,1.5) & 0.93 (0.84,1.03) & 1.08 (0.98,1.18) \\ 
  19 & 0.96 (0.73,1.2) & 0.96 (0.72,1.2) & 0.96 (0.82,1.1) & 0.99 (0.85,1.13) \\ 
  20 & 0.73 (0.53,0.93) & 0.94 (0.75,1.13) & 1.15 (0.95,1.37) & 1.01 (0.8,1.21) \\ 
  21 & 1.01 (0.74,1.29) & 0.94 (0.66,1.21) & 1.08 (0.97,1.19) & 1.02 (0.91,1.13) \\ 
  22 & 1.21 (0.87,1.53) & 1.21 (0.89,1.53) & 1.02 (0.87,1.16) & 1.02 (0.87,1.17) \\ 
  23 & 0.96 (0.7,1.23) & 1.29 (1.02,1.56) & 0.95 (0.84,1.06) & 1.01 (0.91,1.12) \\ 
  24 & 1.16 (0.79,1.55) & 0.69 (0.33,1.06) & 1.08 (0.98,1.18) & 1.11 (1.01,1.2) \\ 
  25 & 1.14 (0.8,1.47) & 0.96 (0.63,1.29) & 0.8 (0.46,1.13) & 0.83 (0.47,1.19) \\ 
   \hline
\end{tabular}
\caption{For $K=25$ partitions, we report parameter estimates for $\beta_{1}$ and $\beta_{2}$ for the simulated count and binary datasets. This includes the posterior mean and 95\% credible intervals for each partition.}
\end{table}

\begin{table}[ht]
\centering
\begin{tabular}{ccccc}
  \hline
  &  \multicolumn{2}{c}{Count Data} &  \multicolumn{2}{c}{Binary Data} \\
 Partition & $\beta_{1}$ & $\beta_{2}$ & $\beta_{1}$ & $\beta_{2}$ \\ 
  \hline
1 & 0.91 (0.68,1.12) & 0.99 (0.77,1.21) & 0.99 (0.94,1.05) & 1.01 (0.95,1.06) \\ 
  2 & 1.81 (0.68,2.88) & 1.22 (0.15,2.28) & 0.99 (0.74,1.24) & 0.95 (0.69,1.2) \\ 
  3 & 1.07 (0.71,1.43) & 0.95 (0.58,1.3) & 0.98 (0.9,1.06) & 1 (0.92,1.08) \\ 
  4 & 0.95 (0.83,1.08) & 1.08 (0.95,1.21) & 0.93 (0.74,1.12) & 0.91 (0.72,1.11) \\ 
  5 & 0.94 (0.64,1.25) & 0.96 (0.64,1.27) & 1.05 (0.93,1.19) & 1.03 (0.91,1.16) \\ 
  6 & 1.19 (0.85,1.53) & 1.03 (0.68,1.36) & 1 (0.84,1.17) & 1.04 (0.88,1.2) \\ 
  7 & 1 (0.69,1.29) & 0.95 (0.65,1.24) & 1 (0.88,1.11) & 1.03 (0.91,1.15) \\ 
  8 & 0.94 (0.72,1.16) & 0.92 (0.7,1.14) & 0.91 (0.78,1.03) & 0.87 (0.75,1) \\ 
  9 & 0.99 (0.77,1.2) & 0.91 (0.7,1.12) & 1.03 (0.82,1.22) & 1.08 (0.87,1.28) \\ 
  10 & 0.84 (-0.08,1.73) & 0.26 (-0.59,1.15) & 0.94 (0.8,1.09) & 0.91 (0.77,1.06) \\ 
  11 & 0.83 (0.45,1.23) & 1.45 (1.05,1.85) & 1 (0.85,1.15) & 1.12 (0.97,1.27) \\ 
  12 & 1.02 (0.8,1.25) & 1.03 (0.8,1.25) & 1.02 (0.94,1.08) & 1.02 (0.95,1.09) \\ 
  13 & 1.01 (0.72,1.29) & 1.02 (0.73,1.3) & 1.04 (0.88,1.19) & 0.96 (0.81,1.12) \\ 
  14 & 0.89 (0.27,1.48) & 0.54 (-0.04,1.17) & 1.01 (0.84,1.18) & 1.11 (0.94,1.28) \\ 
  15 & 0.79 (0.35,1.24) & 0.97 (0.54,1.41) & 0.97 (0.87,1.08) & 1 (0.89,1.1) \\ 
  16 & 1.09 (0.54,1.59) & 1.32 (0.81,1.86) & 1.12 (0.92,1.32) & 1.01 (0.82,1.21) \\ 
  17 & 1.12 (0.94,1.32) & 1.1 (0.91,1.3) & 1.03 (0.88,1.18) & 1.02 (0.87,1.17) \\ 
  18 & 1.08 (0.65,1.5) & 1.06 (0.63,1.5) & 1.1 (0.91,1.28) & 0.87 (0.7,1.05) \\ 
  19 & 1.02 (0.76,1.29) & 1.08 (0.81,1.35) & 1.06 (0.67,1.46) & 0.98 (0.57,1.39) \\ 
  20 & 0.9 (0.72,1.09) & 1.17 (0.99,1.36) & 0.95 (0.78,1.11) & 1.04 (0.87,1.2) \\ 
  21 & 1.06 (0.8,1.32) & 0.97 (0.72,1.24) & 1.02 (0.92,1.12) & 0.99 (0.89,1.09) \\ 
  22 & 0.96 (0.48,1.48) & 0.52 (0.03,1.03) & 1.19 (0.95,1.42) & 1.13 (0.9,1.38) \\ 
  23 & 0.79 (0.38,1.2) & 1.1 (0.67,1.51) & 0.94 (0.71,1.18) & 0.93 (0.7,1.15) \\ 
  24 & 0.83 (0.41,1.26) & 1.06 (0.64,1.51) & 0.93 (0.84,1.03) & 1.08 (0.98,1.18) \\ 
  25 & 0.96 (0.72,1.2) & 0.96 (0.72,1.2) & 0.96 (0.82,1.1) & 0.99 (0.86,1.13) \\ 
  26 & 0.67 (0.42,0.92) & 0.88 (0.62,1.13) & 0.98 (0.78,1.19) & 1.1 (0.89,1.31) \\ 
  27 & 0.82 (0.52,1.16) & 1.04 (0.72,1.35) & 1.18 (0.97,1.39) & 1 (0.79,1.2) \\ 
  28 & 1.04 (0.79,1.28) & 0.99 (0.76,1.25) & 1.02 (0.9,1.13) & 1.08 (0.96,1.19) \\ 
  29 & 1.01 (0.75,1.29) & 0.94 (0.66,1.2) & 1.05 (0.89,1.22) & 1.04 (0.88,1.2) \\ 
  30 & 1.07 (0.66,1.47) & 0.81 (0.4,1.21) & 1.08 (0.97,1.19) & 1.02 (0.91,1.13) \\ 
  31 & 1.21 (0.89,1.54) & 1.19 (0.88,1.53) & 0.89 (0.62,1.14) & 1.2 (0.95,1.47) \\ 
  32 & 1.08 (0.77,1.38) & 1.11 (0.81,1.42) & 1.02 (0.87,1.16) & 1.02 (0.87,1.17) \\ 
  33 & 0.97 (0.69,1.23) & 1.29 (1.01,1.56) & 1 (0.87,1.13) & 1.04 (0.91,1.18) \\ 
  34 & 1.16 (0.79,1.55) & 0.68 (0.31,1.05) & 0.85 (0.66,1.03) & 0.96 (0.78,1.13) \\ 
  35 & 1.14 (0.76,1.52) & 0.82 (0.44,1.19) & 1.08 (0.98,1.18) & 1.11 (1.01,1.2) \\ 
  36 & 1.13 (0.45,1.88) & 1.42 (0.74,2.13) & 0.8 (0.48,1.14) & 0.83 (0.47,1.18) \\ 
   \hline
\end{tabular}
\caption{For $K=36$ partitions, we report parameter estimates for $\beta_{1}$ and $\beta_{2}$ for the simulated count and binary datasets. This includes the posterior mean and 95\% credible intervals for each partition.}
\end{table}


\begin{figure}\label{ModisBeta}
\begin{center}
\includegraphics[ scale = 0.7]{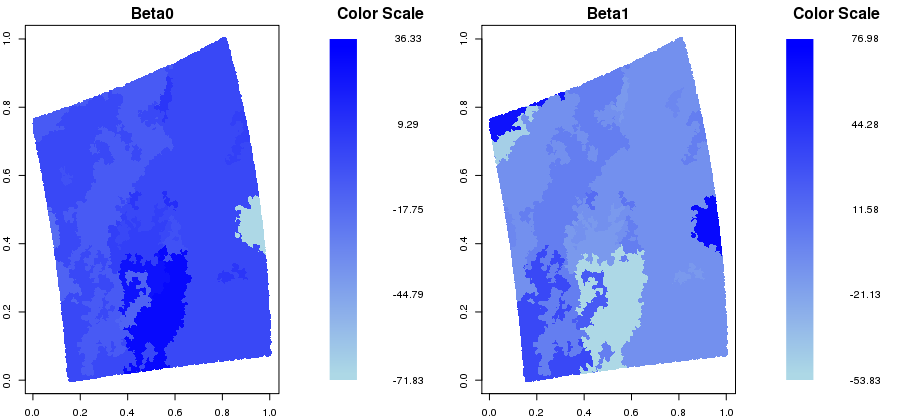}
\end{center}
\caption[]{Map of the spatially-varying $\beta_{0}$ parameter (left) and $\beta_{1}$ parameter (right) for the MODIS cloud mask dataset.)}
\end{figure}

\clearpage

\bibliographystyle{apalike}
\bibliography{Reference}

\begin{thebibliography}{}

\bibitem[Banerjee et~al., 2013]{Banerjee2013}
Banerjee, A., Dunson, D.~B., and Tokdar, S.~T. (2013).
\newblock Efficient {G}aussian process regression for large datasets.
\newblock {\em Biometrika}, 100(1):75--89.

\bibitem[Banerjee and Gelfand, 2006]{banerjee2006bayesian}
Banerjee, S. and Gelfand, A.~E. (2006).
\newblock Bayesian wombling: Curvilinear gradient assessment under spatial
  process models.
\newblock {\em Journal of the American Statistical Association},
  101(476):1487--1501.

\bibitem[Banerjee et~al., 2008]{banerjee2008gaussian}
Banerjee, S., Gelfand, A.~E., Finley, A.~O., and Sang, H. (2008).
\newblock Gaussian predictive process models for large spatial data sets.
\newblock {\em Journal of the Royal Statistical Society: Series B (Statistical
  Methodology)}, 70(4):825--848.

\bibitem[Bradley et~al., 2016]{bradley2016comparison}
Bradley, J.~R., Cressie, N., Shi, T., et~al. (2016).
\newblock A comparison of spatial predictors when datasets could be very large.
\newblock {\em Statistics Surveys}, 10:100--131.

\bibitem[Bradley et~al., 2019]{bradley2019bayesian}
Bradley, J.~R., Holan, S.~H., and Wikle, C.~K. (2019).
\newblock Bayesian hierarchical models with conjugate full-conditional
  distributions for dependent data from the natural exponential family.
\newblock {\em Journal of the American Statistical Association}, 0(ja):1--29.

\bibitem[Christensen et~al., 2006]{Christensen2006robust}
Christensen, O.~F., Roberts, G.~O., and Sk{\"o}ld, M. (2006).
\newblock Robust {M}arkov chain {M}onte {C}arlo methods for spatial generalized
  linear mixed models.
\newblock {\em Journal of Computational and Graphical Statistics}, 15(1):1--17.

\bibitem[Cressie, 2015]{cressie2015statistics}
Cressie, N. (2015).
\newblock {\em Statistics for spatial data}.
\newblock John Wiley \& Sons.

\bibitem[Cressie and Johannesson, 2008]{cressie2008fixed}
Cressie, N. and Johannesson, G. (2008).
\newblock Fixed rank kriging for very large spatial data sets.
\newblock {\em Journal of the Royal Statistical Society: Series B (Statistical
  Methodology)}, 70(1):209--226.

\bibitem[{de Valpine} et~al., 2017]{nimble2017}
{de Valpine}, P., Turek, D., Paciorek, C., Anderson-Bergman, C., {Temple Lang},
  D., and Bodik, R. (2017).
\newblock Programming with models: writing statistical algorithms for general
  model structures with {NIMBLE}.
\newblock {\em Journal of Computational and Graphical Statistics}, 26:403--413.

\bibitem[Diggle et~al., 1998]{diggle1998model}
Diggle, P.~J., Tawn, J., and Moyeed, R. (1998).
\newblock Model-based geostatistics.
\newblock {\em Journal of the Royal Statistical Society: Series C (Applied
  Statistics)}, 47(3):299--350.

\bibitem[Ejigu et~al., 2020]{ejigu2020geostatistical}
Ejigu, B.~A., Wencheko, E., Moraga, P., and Giorgi, E. (2020).
\newblock Geostatistical methods for modelling non-stationary patterns in
  disease risk.
\newblock {\em Spatial Statistics}, 35:100397.

\bibitem[Friedman et~al., 2010a]{friedman2010regularization}
Friedman, J., Hastie, T., and Tibshirani, R. (2010a).
\newblock Regularization paths for generalized linear models via coordinate
  descent.
\newblock {\em Journal of statistical software}, 33(1):1.

\bibitem[Friedman et~al., 2010b]{glmnet2010}
Friedman, J., Hastie, T., and Tibshirani, R. (2010b).
\newblock Regularization paths for generalized linear models via coordinate
  descent.
\newblock {\em Journal of Statistical Software}, 33(1):1--22.

\bibitem[Fuentes, 2001]{fuentes2001high}
Fuentes, M. (2001).
\newblock A high frequency kriging approach for non-stationary environmental
  processes.
\newblock {\em Environmetrics: The official journal of the International
  Environmetrics Society}, 12(5):469--483.

\bibitem[Fuentes, 2002]{fuentes2002interpolation}
Fuentes, M. (2002).
\newblock Interpolation of nonstationary air pollution processes: a spatial
  spectral approach.
\newblock {\em Statistical Modelling}, 2(4):281--298.

\bibitem[Gopal et~al., 2019]{gopal2019characterizing}
Gopal, S., Ma, Y., Xin, C., Pitts, J., and Were, L. (2019).
\newblock Characterizing the spatial determinants and prevention of malaria in
  {K}enya.
\newblock {\em International journal of environmental research and public
  health}, 16(24):5078.

\bibitem[Griffith, 2003]{griffith2003spatial}
Griffith, D.~A. (2003).
\newblock Spatial filtering.
\newblock In {\em Spatial Autocorrelation and Spatial Filtering}, pages
  91--130. Springer.

\bibitem[Guan and Haran, 2018]{guan2018computationally}
Guan, Y. and Haran, M. (2018).
\newblock A computationally efficient projection-based approach for spatial
  generalized linear mixed models.
\newblock {\em Journal of Computational and Graphical Statistics},
  27(4):701--714.

\bibitem[Guhaniyogi and Banerjee, 2018]{guhaniyogi2018meta}
Guhaniyogi, R. and Banerjee, S. (2018).
\newblock Meta-kriging: Scalable {B}ayesian modeling and inference for massive
  spatial datasets.
\newblock {\em Technometrics}, 60(4):430--444.

\bibitem[Haran et~al., 2003]{haran2003accelerating}
Haran, M., Hodges, J.~S., and Carlin, B.~P. (2003).
\newblock Accelerating computation in markov random field models for spatial
  data via structured mcmc.
\newblock {\em Journal of Computational and Graphical Statistics},
  12(2):249--264.

\bibitem[Heaton et~al., 2017]{heaton2017nonstationary}
Heaton, M.~J., Christensen, W.~F., and Terres, M.~A. (2017).
\newblock Nonstationary gaussian process models using spatial hierarchical
  clustering from finite differences.
\newblock {\em Technometrics}, 59(1):93--101.

\bibitem[Hefley et~al., 2017]{hefley2017basis}
Hefley, T.~J., Broms, K.~M., Brost, B.~M., Buderman, F.~E., Kay, S.~L., Scharf,
  H.~R., Tipton, J.~R., Williams, P.~J., and Hooten, M.~B. (2017).
\newblock The basis function approach for modeling autocorrelation in
  ecological data.
\newblock {\em Ecology}, 98(3):632--646.

\bibitem[Higdon, 1998]{higdon1998process}
Higdon, D. (1998).
\newblock A process-convolution approach to modelling temperatures in the
  {N}orth {A}tlantic {O}cean.
\newblock {\em Environmental and Ecological Statistics}, 5(2):173--190.

\bibitem[Higdon et~al., 2008]{higdon2008computer}
Higdon, D., Gattiker, J., Williams, B., and Rightley, M. (2008).
\newblock Computer model calibration using high-dimensional output.
\newblock {\em Journal of the American Statistical Association},
  103(482):570--583.

\bibitem[Holland et~al., 1999]{holland1999spatial}
Holland, D.~M., Saltzman, N., Cox, L.~H., and Nychka, D. (1999).
\newblock Spatial prediction of sulfur dioxide in the eastern {U}nited
  {S}tates.
\newblock In {\em geoENV II—Geostatistics for environmental applications},
  pages 65--76. Springer.

\bibitem[Hughes and Haran, 2013]{hughes2013dimension}
Hughes, J. and Haran, M. (2013).
\newblock Dimension reduction and alleviation of confounding for spatial
  generalized linear mixed models.
\newblock {\em Journal of the Royal Statistical Society: Series B (Statistical
  Methodology)}, 75(1):139--159.

\bibitem[ICF, 2020]{ICF2017}
ICF (2004-2017 (Accessed July, 1, 2020)).
\newblock Demographic and health surveys (various) [datasets].
\newblock Funded by USAID. Data retrieved from ,
  \url{http://dhsprogram.com/data/available-datasets.cfm}.

\bibitem[Katzfuss, 2013]{katzfuss2013bayesian}
Katzfuss, M. (2013).
\newblock Bayesian nonstationary spatial modeling for very large datasets.
\newblock {\em Environmetrics}, 24(3):189--200.

\bibitem[Katzfuss, 2017]{katzfuss2017multi}
Katzfuss, M. (2017).
\newblock A multi-resolution approximation for massive spatial datasets.
\newblock {\em Journal of the American Statistical Association},
  112(517):201--214.

\bibitem[Kim et~al., 2005]{kim2005analyzing}
Kim, H.-M., Mallick, B.~K., and Holmes, C. (2005).
\newblock Analyzing nonstationary spatial data using piecewise gaussian
  processes.
\newblock {\em Journal of the American Statistical Association},
  100(470):653--668.

\bibitem[Kleiber and Nychka, 2012]{kleiber2012nonstationary}
Kleiber, W. and Nychka, D. (2012).
\newblock Nonstationary modeling for multivariate spatial processes.
\newblock {\em Journal of Multivariate Analysis}, 112:76--91.

\bibitem[Lee and Haran, 2019]{lee2019picar}
Lee, B.~S. and Haran, M. (2019).
\newblock Picar: An efficient extendable approach for fitting hierarchical
  spatial models.
\newblock {\em arXiv preprint arXiv:1912.02382}.

\bibitem[Minsker et~al., 2017]{minsker2017robust}
Minsker, S., Srivastava, S., Lin, L., and Dunson, D.~B. (2017).
\newblock Robust and scalable {B}ayes via a median of subset posterior
  measures.
\newblock {\em The Journal of Machine Learning Research}, 18(1):4488--4527.

\bibitem[Nychka et~al., 2015]{nychka2015multiresolution}
Nychka, D., Bandyopadhyay, S., Hammerling, D., Lindgren, F., and Sain, S.
  (2015).
\newblock A multiresolution {G}aussian process model for the analysis of large
  spatial datasets.
\newblock {\em Journal of Computational and Graphical Statistics},
  24(2):579--599.

\bibitem[Nychka et~al., 2002]{nychka2002multiresolution}
Nychka, D., Wikle, C., and Royle, J.~A. (2002).
\newblock Multiresolution models for nonstationary spatial covariance
  functions.
\newblock {\em Statistical Modelling}, 2(4):315--331.

\bibitem[O'Hara et~al., 2009]{o2009review}
O'Hara, R.~B., Sillanp{\"a}{\"a}, M.~J., et~al. (2009).
\newblock A review of {B}ayesian variable selection methods: what, how and
  which.
\newblock {\em Bayesian analysis}, 4(1):85--117.

\bibitem[Paciorek and Schervish, 2006]{paciorek2006spatial}
Paciorek, C.~J. and Schervish, M.~J. (2006).
\newblock Spatial modelling using a new class of nonstationary covariance
  functions.
\newblock {\em Environmetrics: The official journal of the International
  Environmetrics Society}, 17(5):483--506.

\bibitem[Risser and Calder, 2015]{risser2015local}
Risser, M.~D. and Calder, C.~A. (2015).
\newblock Local likelihood estimation for covariance functions with
  spatially-varying parameters: the convospat package for r.
\newblock {\em arXiv preprint arXiv:1507.08613}.

\bibitem[Royle and Wikle, 2005]{royle2005efficient}
Royle, J.~A. and Wikle, C.~K. (2005).
\newblock Efficient statistical mapping of avian count data.
\newblock {\em Environmental and Ecological Statistics}, 12(2):225--243.

\bibitem[Rue et~al., 2009]{rue2009approximate}
Rue, H., Martino, S., and Chopin, N. (2009).
\newblock Approximate bayesian inference for latent gaussian models by using
  integrated nested laplace approximations.
\newblock {\em Journal of the royal statistical society: Series b (statistical
  methodology)}, 71(2):319--392.

\bibitem[Sampson, 2010]{sampson2010constructions}
Sampson, P. (2010).
\newblock Constructions for nonstationary spatial processes in gelfand ae,
  diggle pj, fuentes m, and guttorp p (ed.), the handbook of spatial
  statistics, chapter 9.

\bibitem[Scott et~al., 2016]{scott2016bayes}
Scott, S.~L., Blocker, A.~W., Bonassi, F.~V., Chipman, H.~A., George, E.~I.,
  and McCulloch, R.~E. (2016).
\newblock Bayes and big data: The consensus {M}onte {C}arlo algorithm.
\newblock {\em International Journal of Management Science and Engineering
  Management}, 11(2):78--88.

\bibitem[Sengupta and Cressie, 2013]{sengupta2013hierarchical}
Sengupta, A. and Cressie, N. (2013).
\newblock Hierarchical statistical modeling of big spatial datasets using the
  exponential family of distributions.
\newblock {\em Spatial Statistics}, 4:14--44.

\bibitem[Shaby and Wells, 2010]{shaby2010exploring}
Shaby, B. and Wells, M.~T. (2010).
\newblock Exploring an adaptive metropolis algorithm.
\newblock {\em Currently under review}, 1(1):17.

\bibitem[Shi and Cressie, 2007]{shi2007global}
Shi, T. and Cressie, N. (2007).
\newblock Global statistical analysis of misr aerosol data: a massive data
  product from nasa's terra satellite.
\newblock {\em Environmetrics: The official journal of the International
  Environmetrics Society}, 18(7):665--680.

\bibitem[Tibshirani, 1996]{tibshirani1996regression}
Tibshirani, R. (1996).
\newblock Regression shrinkage and selection via the lasso.
\newblock {\em Journal of the Royal Statistical Society: Series B
  (Methodological)}, 58(1):267--288.

\bibitem[Zilber and Katzfuss, 2020]{zilber2020vecchia}
Zilber, D. and Katzfuss, M. (2020).
\newblock Vecchia-{L}aplace approximations of generalized {G}aussian processes
  for big non-{G}aussian spatial data.
\newblock {\em Computational Statistics \& Data Analysis}, page 107081.

\end{thebibliography}
\end{document}